\documentclass[]{interact}

\usepackage{epstopdf}
\usepackage{subcaption}
\usepackage{booktabs}
\usepackage{array}
\usepackage{graphicx}
\usepackage{hyperref}
\usepackage[export]{adjustbox}
\usepackage[normalem]{ulem}
\usepackage{xcolor}

\usepackage[longnamesfirst,sort]{natbib}

\newcommand{\remove}[1]{}
\newcommand{\add}[1]{#1}

\newif\ifanonymous
\anonymousfalse     

\bibpunct[, ]{(}{)}{;}{a}{,}{,}
\renewcommand\bibfont{\fontsize{10}{12}\selectfont}

\theoremstyle{plain}

\theoremstyle{definition}

\theoremstyle{remark}

\begin{document}

\articletype{RESEARCH ARTICLE}

\title{Supporting Researcher Control in AI-Assisted Social Media Analysis: A Mixed-Methods Study of AI4OSN}

\author{
\name{Qile Wang\textsuperscript{a*}, Abhishek Karwankar\textsuperscript{a}, Shukria Muhammadi\textsuperscript{a}, Srikanth Kumar Perumalla\textsuperscript{a}, Abhin Suresh\textsuperscript{a}, Sunita Chandrasekaran\textsuperscript{a}, Kenneth E.~Barner\textsuperscript{b}, and Matthew Louis Mauriello\textsuperscript{a}\thanks{CONTACT Qile Wang. Email: kylewang@udel.edu}}
\affil{\textsuperscript{a}Department of Computer \& Information Sciences, University of Delaware, Newark, DE, USA}
\affil{\textsuperscript{b}Department of Electrical and Computer Engineering, University of Delaware, Newark, DE, USA}
}


\maketitle

\begin{abstract}
High-quality annotated datasets are critical for advancing computational social science (CSS) research, yet collecting, curating, and analyzing large-scale data from dynamic sources, such as online social networks (OSNs), remain persistent bottlenecks. Researchers must also translate substantive questions into queries, relevance criteria, and defensible interpretations. Large language models (LLMs) can assist but can also obscure judgments, potentially reducing researcher oversight and critical evaluation. We present AI4OSN, a mixed-initiative platform that keeps researchers in control of data collection, relevance labeling, and sensemaking while making AI assistance available as an inspectable advisory resource. In a systematic user study (N=18) using X/Twitter data, participants reported above-benchmark perceived usability (mean SUS score of 79.7), \remove{increased confidence in LLM-assisted annotation and analysis tasks, and low cognitive workload during AI-supported sensemaking.} \add{increased self-reported confidence in data collection, labeling, and analysis.} Notably, novice participants showed descriptively greater alignment and higher reported agreement with AI output than experienced participants, underscoring the need to support differentiated verification practices across expertise levels. \add{These group comparisons are exploratory and motivate controlled follow-up studies of how AI assistance shapes confidence and annotation quality.} Overall, this work contributes both a human-centered system for social data analysis and empirical insights into how AI-assisted interfaces shape researcher behavior, trust, and decision-making. Our findings highlight design opportunities and tensions in building accessible, transparent, and agentic AI systems for social media and web research.

\end{abstract}

\begin{keywords}
Human-AI collaboration; large language models; social media research; mixed-initiative systems; trust calibration, usability research
\end{keywords}

\section{Introduction}

\begin{figure}[!ht]
    \centering
    \includegraphics[width=0.7\linewidth]{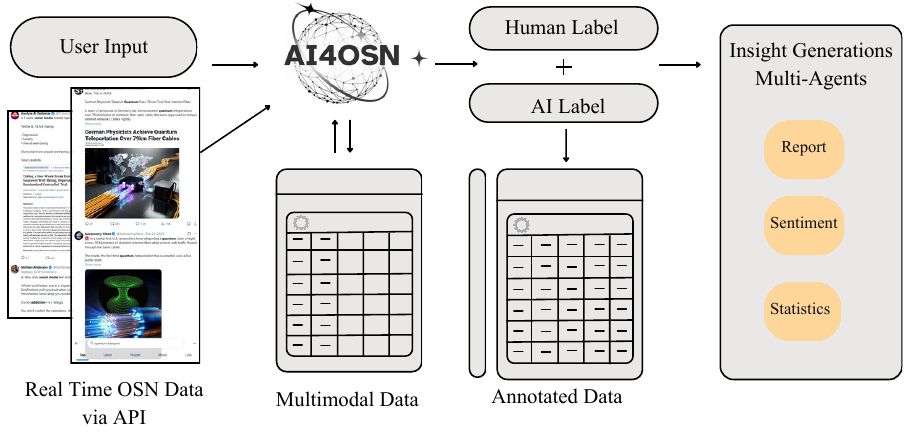}
    \caption{An overview of the AI4OSN design, which integrates an LLM-powered multi-agent into mixed-initiative workflows for data collection, annotation, and post-analysis. The pipeline runs from left to right, taking user-input keywords to fetch real-time OSN data via API calls. The data is then saved into each user's Firebase account. The saved data can be labeled either manually by users or by AI agents using user-defined prompts. The final stage of the pipeline includes a multi-agent system to extract summaries, general sentiment, and overall statistics from the annotated data.}
    \label{fig:overall_figure}
\end{figure}

Online social networks (OSNs) are widely used to study public opinion and social behavior across various domains, including social science \citep{sapountzi2018social}, business \citep{ferreira2009effect, fuciu2013marketing},  health \citep{alshaikh2014social, benkaouz2015towards, blanco2016online}, transportation \citep{rashidi2017exploring}, education \citep{hamid2015understanding}, and crisis \citep{10.1145/3577213}. Despite their importance, accessing and analyzing OSN data remains difficult for many researchers \citep{stieglitz2018social, wang2025lata}. Existing workflows often depend on APIs, scripting, and manual annotation, which require substantial technical expertise and sustained effort. These barriers limit participation in Computational Social Science (CSS) research and disproportionately affect researchers with limited programming backgrounds. At the same time, prior work on data cascades shows how early data quality issues can propagate and undermine downstream analysis \citep{sambasivan2021everyone}, underscoring the need for high-quality labeled data at the outset. Producing such data is therefore not only a technical problem but also an interaction design challenge. Researchers must make sense of noisy and rapidly evolving content while maintaining transparency and methodological rigor \citep{sapountzi2018social}. However, many existing tools emphasize full automation over supporting user understanding and control, leaving a critical gap for human-centered approaches~\citep{amershi2019guidelines, Bansal2021Whole, Ben2020Human}.

LLMs create opportunities to support annotation, summarization, and sensemaking at scale, but they can also introduce bias, ambiguity, and inappropriate reliance \citep{thapa2025large, ashery2025emergent, Mei2026Passing, wang2026explanation}. In human-AI decision-making, people need opportunities to interrogate, correct, and contextualize AI output rather than merely receive it \citep{amershi2019guidelines, bucinca2021trust}. Research in data analytics similarly shows that trust is shaped by both perceived system behavior and users' opportunities to verify its recommendations \citep{wang2019human, tutul2024trust, bach2024trust}. These concerns are especially consequential in social media research, where a seemingly simple relevance judgment may embed substantive, contextual, and ethical choices.

With these concerns in mind, we develop AI4OSN (AI-assisted platform for online social networks) as a mixed-initiative research platform designed around three requirements: (R1) preserve researchers' ownership of queries and relevance criteria; (R2) make AI-mediated transformations inspectable and revisable; and (R3) support calibration by exposing AI suggestions as advisory inputs. The system instantiates these requirements through project-based workspaces, manual-first labeling, optional prompt-configured AI suggestions, and source-linked analysis outputs (Figure~\ref{fig:overall_figure}). Building on earlier social media research systems (e.g.,  \citep{byun2012automated, benitez2017web, mauriello2018smidgen, mauriello2018large}), we developed our system to evaluate the following questions:

\begin{itemize}

\item \textbf{RQ1}: How do novice and experienced researchers perceive and interact with a researcher-controlled, AI-assisted workflow for collecting, labeling, and interpreting social media data?

\item \textbf{RQ2}: How do users adopt, verify, and evaluate AI-generated suggestions while applying their own relevance criteria?

\item  \textbf{RQ3}: What novel interface and workflow supports do users identify as necessary for responsible reliance on AI-assisted analysis?

\end{itemize}

We present an AI4OSN system that integrates LLM assistance into social media research workflows while preserving researcher agency over queries, labeling, and interpretation. To understand how researchers perceive and use such AI-assisted workflows, we conducted a mixed-methods study with 18 participants. Rather than comparing AI and human accuracy, the study characterizes perceived usability, revision behavior, and accounts of trust and verification. Our contributions are:
\begin{itemize}
    \item We introduce AI4OSN, demonstrating a design approach to human-centered, researcher-controlled AI assistance for social media research

    \item We establish a set of design requirements for researcher-controlled AI assistance in social media research, grounded in human-computer interaction (HCI) principles regarding human control, verification, and reliance.

    \item empirical findings showing how participants treated AI suggestions as contestable evidence, which motivate design implications for prompt support, provenance, uncertainty communication, and human revision.
\end{itemize}

\section{Related Work}
This section reviews prior work on social media and web-based research, particularly studies that focus on large-scale data collection, filtering, and analysis, as well as the challenges of data quality and bias. We also review research on human–AI collaboration that examines AI support for human decision making, including trust, transparency, and oversight.

\subsection{Social Media and Web Research}

OSNs have become a central data source for computational social science (CSS), enabling large-scale analysis of public opinion, behavior, and social dynamics. Prior work has leveraged platforms such as Twitter, Reddit, and Facebook to study misinformation \citep{saxena2023statistical, guo2020future, 10.1145/3688598}, bias \citep{carrascosa2015quantifying, kulshrestha2017quantifying}, social trends \citep{allcott2019trends, 10.1145/3665450}, and a wide range of related domains. An extensive review of more than 2,300 studies summarizes common pipelines and application domains. It identifies open challenges related to temporal dynamics, causal inference, and the gap between academic methods and commercial practice \citep{rodriguez2023review}. This review further highlights core challenges, including noisy data quality, over-reliance on traditional techniques (lexicons, tokens, Bayesian methods, bag-of-words), trade-offs between scalability and interpretability, and limited support for end-to-end research workflows.

\subsubsection{OSN Collection}
Beyond existing curated OSN datasets, several platforms and frameworks have been developed to support OSN collection by automating parts of the data collection and labeling process. Toolkits such as Twarc\footnote{\url{https://twarc-project.readthedocs.io/}} facilitate large-scale data collection from X.com (previously Twitter) APIs. In contrast, Pushshift\footnote{\url{https://pushshift.io/}} has been widely used to archive and retrieve Reddit data for research purposes. In addition, annotation pipelines and workflow-based systems support downstream labeling and preprocessing of social media data \citep{bernstein2011crowdsourcing}. Although these tools exist, they often require technical expertise and have limited support for integrated analysis and human-centered workflows.

\subsubsection{OSN Analysis \& Platform}
In addition to the technical details of data collection, researchers commonly employ a range of analytic techniques, including topic modeling, time series analysis, sentiment analysis, network analysis, and clustering, to examine patterns in public opinion, behavior, and social interaction \citep{ghosh2025intersection}. Among these methods, sentiment analysis remains one of the most widely used techniques. Surveys of aspect-based sentiment analysis highlight persistent difficulties in extracting meaningful aspects and aligning them with their correct polarity in complex, noisy language \citep{nazir2020issues}. 
Studies on contextual sentiment analysis show that lexicon-based approaches struggle when sentiment depends on genre and situational context \citep{muhammad2016contextual}.
Several platforms are currently available, such as the PennMAP\footnote{\url{https://css.seas.upenn.edu/project/penn-map/}}, GDELT\footnote{\url{https://www.gdeltproject.org}}, and MediaClound\footnote{\url{https://search.mediacloud.org/}}, to investigate media transparency and accountability. However, these platforms primarily support static data exploration and lack interactive capabilities for human-in-the-loop analysis and real-time data collection. In contrast, our approach enables interactive workflows and supports real-time data collection, allowing for dynamic and scalable human–AI collaboration.

\subsection{Human-AI Collaboration}
Human–AI collaboration has become a central concern in Human-Computer Interaction (HCI) research, particularly as AI systems become more prevalent in analytical and creative domains. Prior work emphasizes explainability, interactivity, and complementarity, in which people and AI systems have distinct roles in jointly producing outcomes \citep{amershi2019guidelines, Bansal2021Whole, Ben2020Human}. Building on foundational work on mixed-initiative systems \citep{Horvitz1999Principles}, recent research emphasizes designs in which both human and AI agents contribute actively to goal achievement as a promising approach for supporting human judgment while leveraging AI capabilities \citep{amershi2019guidelines}. Core design principles include supporting user control, enabling appropriate reliance on AI suggestions, promoting transparency and accountability, and facilitating human-AI alignment on task-critical values.

Recent work has further examined how interaction design can improve trust and understanding in AI-assisted decision-making. In data exploration and visualization, interactive explanations can help users build more accurate mental models of AI behavior \citep{wang2019human}. In high-stakes decision settings, requiring users to reflect before accepting an AI recommendation can reduce over-reliance \citep{bucinca2021trust}. These studies highlight a recurring pattern: transparency alone is insufficient. Rather, users need opportunities for direct engagement with AI capabilities to calibrate reliance appropriately.
Empirical research similarly finds that the effects of explanation depend on the form of explanation and the outcome being measured, reinforcing the need to evaluate explanation features in context rather than assuming that more information automatically produces better human-AI teaming \citep{silva2023xai}.

\subsubsection{Researcher Control and Trust Calibration}

For research workflows, the central design question is not whether an AI system can produce a label or summary, but how its output can be held accountable by the people interpreting it. Human-AI interaction guidance emphasizes maintaining human control, supporting efficient correction, and helping users understand when to rely on system output \citep{amershi2019guidelines}. \add{\citet{marozzo2026ai} argues that AI can act as a bridge across fragmented domains of expertise, but only within broader human, institutional, and technical coordination. The author frames effective use as hybrid sociotechnical agency and concludes that accountable human–AI collaboration is more promising than autonomous substitution. AI4OSN applies this logic to com-
putational social science, where collecting, filtering, and interpreting posts from online
social networks requires moving across technical and substantive expertise.} Research on onboarding for human-AI collaboration further emphasizes the need to communicate a system's capabilities, limits, and the user's role in the workflow \citep{cai2019hello}. Work on cognitive forcing functions shows that interfaces can reduce over-reliance by requiring users to engage critically with AI recommendations \citep{bucinca2021trust}. In data analytics tasks, agreement behavior and self-reported trust can diverge across cases; consequently, designers should treat trust as an evolving calibration problem rather than as a proxy for system quality \citep{tutul2024trust}.

We derive three requirements for AI-assisted social media research: researchers must be able to articulate and revise their own relevance criteria; they must be able to inspect the relationship between AI output and the underlying posts; and they must be able to treat AI suggestions as revisable inputs rather than final decisions. These requirements guide the AI4OSN workflow and our evaluation. Although the platform can link users to associated media, the present study evaluates a text-centered X/Twitter workflow.

\subsubsection{Perceptions of AI-Assisted Tools}

Recent research increasingly examines how AI-generated outputs shape user trust, confidence, and analytical behavior, motivating evaluations that extend beyond AI model accuracy \citep{kalving2024ai}. Users' trust in AI systems is closely tied to how outputs are presented, explained, and contextualized within workflows \citep{hoffman2018metrics, amershi2019guidelines}. Reliance on LLM-generated annotations introduces new risks that can distort downstream analysis. Recent work demonstrates that seemingly reasonable design choices, such as prompt formulation and model selection, can substantially alter statistical conclusions, a phenomenon characterized as ``LLM hacking'' \citep{baumann2025large}. These findings underscore the need for validation, transparency, and human oversight in AI-assisted research workflows.

User perceptions and attitudes toward AI-assisted tools are shaped by factors such as the perceived source of suggestions, semantic clarity, and explainability. For instance, data science practitioners evaluate and trust AI-generated features differently depending on whether outputs are framed as human-produced or AI-produced, with transparency playing a key role in adoption and perceived usefulness \citep{zhu2024towards}. Research on explainable AI further shows that poorly aligned explanations can lead to over-reliance or skepticism rather than calibrated trust \citep{kizilcec2016much}.

Recent qualitative evidence similarly shows that people interact with generative AI differently across phases of a task, valuing it as a benchmark, booster, executor, or amplifier while remaining concerned about superficial engagement \citep{sandhaus2025co}. In social media research, these concerns are compounded by practical barriers such as workflow fragmentation, data access complexity, and trust in automated outputs \citep{wang2025lata}. Rather than treating AI assistance as an independent analytic authority, our study examines how a mixed-initiative workflow can support researchers' own criteria, verification practices, and responsibility for interpretation.

\section{AI4OSN Platform Development}

\textbf{Design rationale.} AI4OSN was designed to support researcher-controlled transitions between three stages of social media inquiry: converting a research interest into a query, mapping that query onto relevance judgments, and distilling a labeled collection into preliminary interpretations. The interface operationalizes the requirements identified above by retaining user-authored queries and labels as the primary record (R1), keeping source posts and configurable AI inputs available for inspection (R2), and presenting AI labels and summaries as suggestions that users may review, revise, or reject (R3). The following description distinguishes implemented capabilities from proposed future extensions.

\begin{figure}[!bht]
    \centering
    \includegraphics[width=0.7\linewidth, frame]{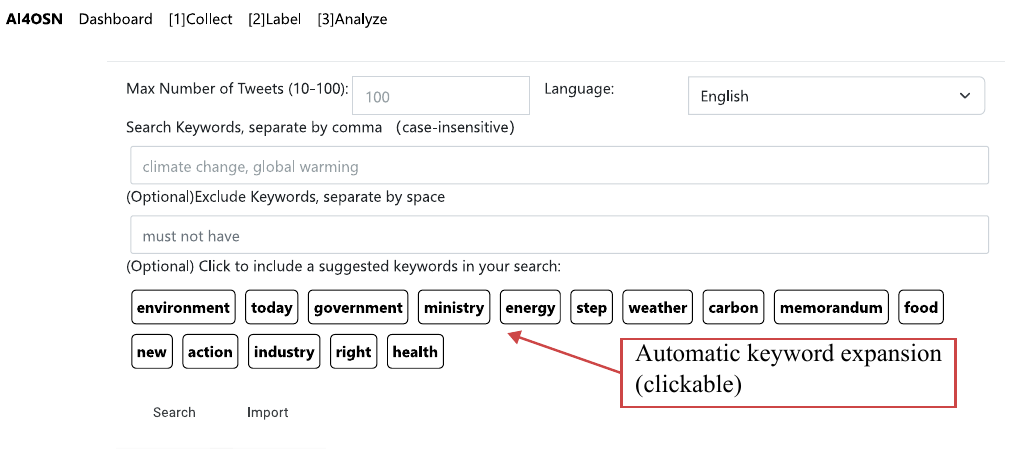}
    \caption{The first component of the platform ingests real-time OSN data. Users can specify multiple keywords, using spaces to perform intersection (AND) searches and commas to perform union (OR) searches. If the user is satisfied with the search results, they can save them to their Firebase account for persistent storage and later use.}
    \label{fig:search_keyword}
\end{figure}

\begin{figure}[!bht]
    \centering
    \includegraphics[width=0.7\linewidth, frame]{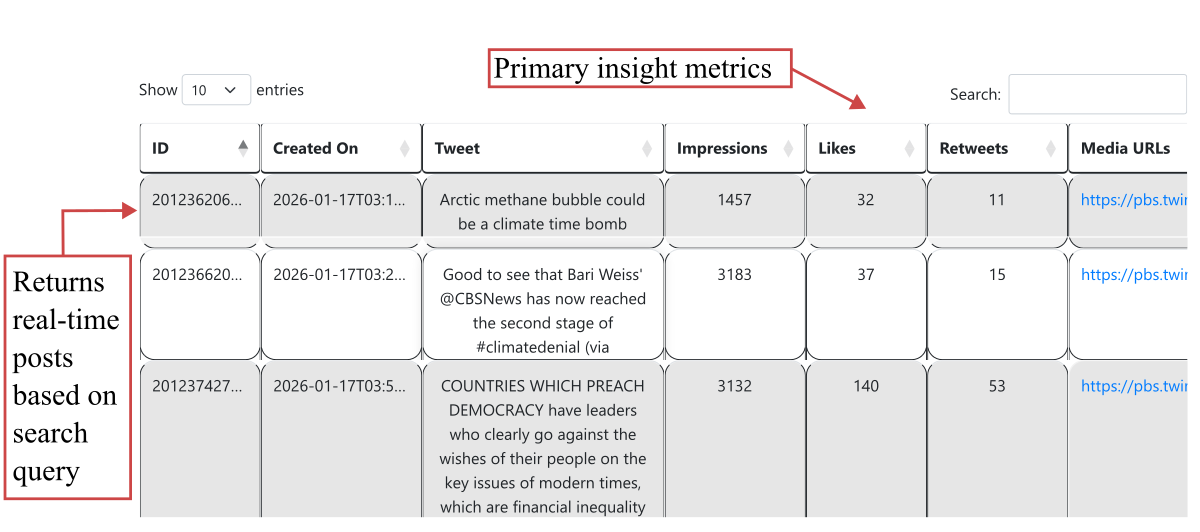}
    \caption{The OSN data returned from the keyword search are displayed in a tabular format, showing key engagement metrics such as impressions, likes, and retweets. Users can sort posts based on any of these metrics. The rightmost column provides links to associated images, allowing users to access additional content details when needed.}
    \label{fig:search_results}
\end{figure}

\begin{figure}[!ht]
    \centering
    \includegraphics[width=0.7\linewidth, frame]{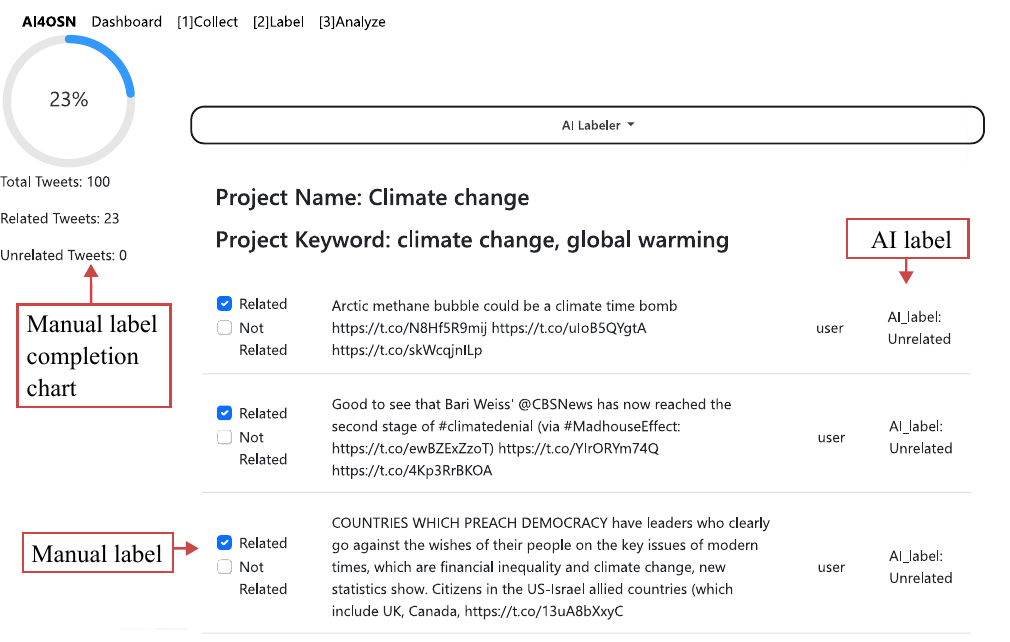}
    \caption{The second component of the platform supports manual labeling of saved search results. Users can currently classify each post as related or unrelated based on its relevance to the search keywords. A label progress chart on the left provides basic statistics and an overview of the manual labeling progress.}
    \label{fig:labeling}
\end{figure}

\begin{figure}[!ht]
    \centering
    \includegraphics[width=0.7\linewidth, frame]{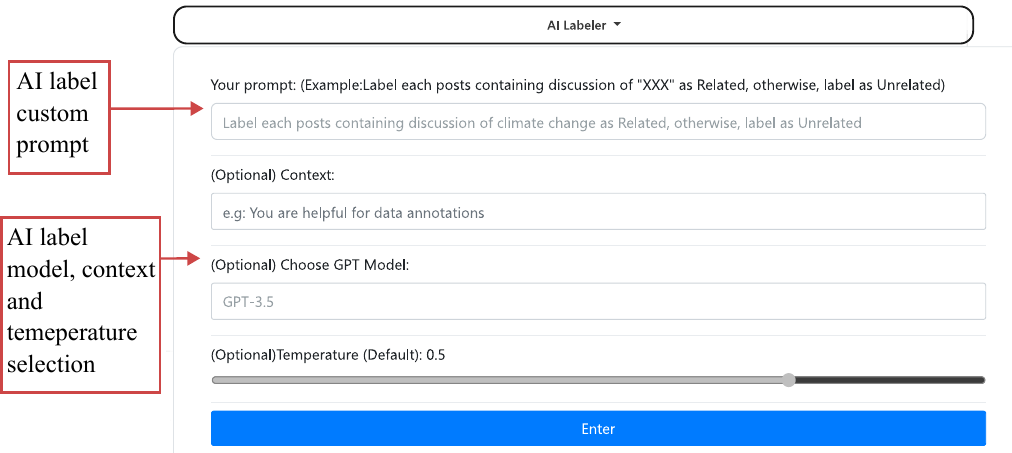}
    \caption{To assist with manual labeling, the platform provides an AI-based labeling option. Users can provide a custom prompt and select the AI model, Gen-AI system context, and model temperature to label search results. The labeled data can then be saved to the user's Firebase account for persistent storage and later use.}
    \label{fig:ailabel_prompt}
\end{figure}

\begin{figure}[!ht]
    \centering
    \includegraphics[width=0.6\linewidth, frame]{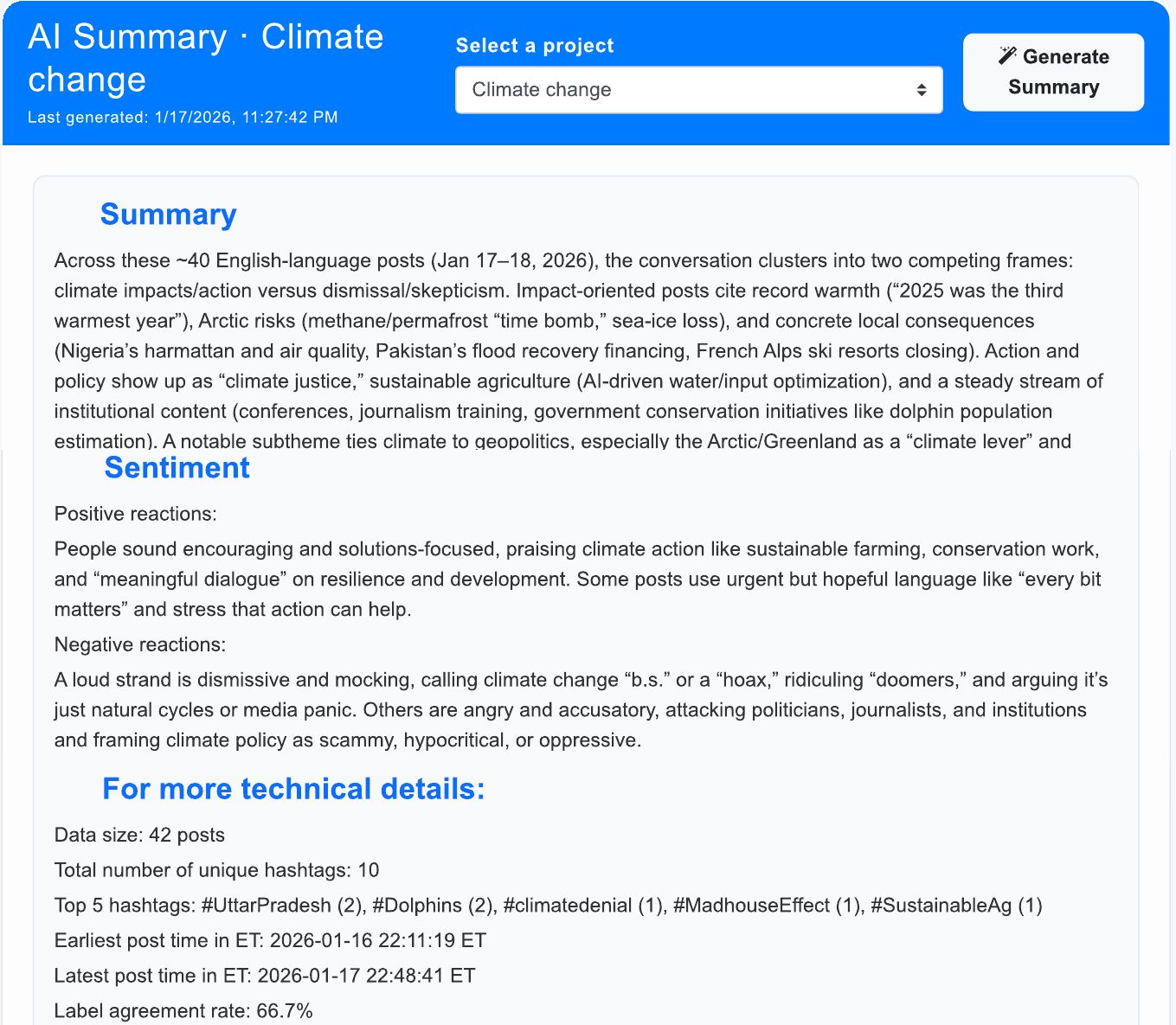}
    \caption{The final component of the platform generates AI-driven summaries of the collected resources. The output is currently divided into three sections: a content summary, sentiment analysis, and statistical information. Dedicated AI agents are assigned to each task to perform these analyses independently.}
    \label{fig:ai_summary}
\end{figure}

AI4OSN is a web-based, mixed-initiative platform for social media research. It integrates data collection, researcher-owned annotation, and early-stage sensemaking into a single project workspace; Figure~\ref{fig:overall_figure} summarizes this workflow.
Users access the system through an account-based login that enables persistent projects and collaboration. The project dashboard allows users to create studies, manage existing work, and invite collaborators. Each project serves as a self-contained workspace for its data, annotations, and analysis results.

First, for each project, users configure data collection via an interactive search interface that supports keyword queries, language filters, and multiple post limits (as shown in Figure \ref{fig:search_keyword}). Automatic keyword expansion provides clickable suggestions that support iterative refinement without technical query syntax. Retrieved posts are displayed with basic metadata and engagement indicators to support early relevance assessment (Figure \ref{fig:search_results}). For this study, we use the real-time X.com/Twitter API.

The labeling stage foregrounds human judgment. Users first apply their own relevance labels to collected posts; each label is associated with its annotator, and progress indicators make the scope of completed work visible (Figure~\ref{fig:labeling}). Users can subsequently invoke an optional AI labeler by providing task instructions and selecting GPT 3.5, GPT 4, or GPT 4 Turbo, together with a system context and temperature setting (Figure~\ref{fig:ailabel_prompt}). AI labels are displayed as suggestions, while human labels remain the primary record used by downstream analysis. The interface supports binary relevance labeling in the study and can support multi-class labels when configured by a project.

The analysis interface applies a LangGraph based workflow\footnote{\url{https://www.langchain.com/langgraph}} only after relevance filtering (Figure~\ref{fig:ai_summary}). Human-assigned relevance labels take precedence, and AI labels are used only where no human label is available. The workflow produces a content summary, a sentiment description, and descriptive statistics. An integrated chatbox supports follow-up questions over the collected data. These outputs are intended for preliminary sensemaking and require researchers to inspect the underlying posts before treating them as research findings.

\section{Usability Study}

Following established practices for usability evaluation in HCI, we conducted a task-based, mixed-method evaluation in which participants completed representative system workflows while we collected complementary behavioral, subjective, and qualitative measures \citep{lazar2017research, bastien2010usability, hornbaek2006current}. This approach enables usability to be examined through both participants' interactions with the system and their reported experiences, rather than relying on a single post-use measure \citep{hornbaek2006current}. Similar task-based and mixed-method protocols have also been employed in recent evaluations of AI-assisted and sensemaking systems \citep{rachatasumrit2021forsense, fan2022human, karwankar2026resonavisvisualizinginteractivemusic}. The goal here was to characterize how participants experienced the workflow, applied their own relevance criteria, and described trust and verification when AI suggestions were introduced. The study received approval from the Institutional Review Board of the affiliated university
\ifanonymous
(Protocol \#\#\#\#\#\#\#).
\else
(Protocol \#2037012-3).
\fi

The overall study design comprised three modules (duration of approximately 45 minutes): (1) a pre-survey, (2) an interface interaction phase, and (3) a post-survey followed by a semi-structured debrief interview. 
During the interface interaction phase, participants completed a sequence of four tasks reflecting the core workflow of \textsc{AI4OSN}: (1) project creation, (2) query configuration, (3) data review and relevance labeling, and (4) AI-assisted analysis and interpretation. Each task was instrumented with system logging, post-task questionnaires, and observational notes to capture both performance and subjective experience. The study concluded with a standardized usability assessment and a semi-structured interview to elicit reflective feedback on usability, trust, and sensemaking.


\begin{table}[tbh]
    \centering
    \footnotesize
    \caption{Participant Demographics ($N = 18$): counts ($n$) and percentages (\%) for age, gender, political affiliation, education, and domain of expertise.}

    \begin{tabular}{@{}lcc@{\hspace{2.5em}}lcc@{}}
    \toprule
    \textbf{Age} & \textbf{$n$} & \textbf{\%} & \textbf{Gender} & \textbf{$n$} & \textbf{\%} \\
    \midrule
    $<24$ & 9 & 50.0 & Male & 8 & 44.4 \\
    25--30 & 4 & 22.2 & Female & 10 & 55.6 \\
    31--36 & 3 & 16.7 & Non-binary & --- & --- \\
    37--42 & 1 & 5.6 & Prefer not to say & --- & --- \\
    $43+$ & 1 & 5.6 & & & \\
    \midrule
    \textbf{Political Affiliation} & \textbf{$n$} & \textbf{\%} & \textbf{Education Completed} & \textbf{$n$} & \textbf{\%} \\
    \midrule
    Democrat & 3 & 16.7 & High school/GED & 2 & 11.1 \\
    Republican & --- & --- & Some College & 3 & 16.7 \\
    Independent & 3 & 16.7 & Bachelor's Degree & 3 & 16.7 \\
    Don't know & 3 & 16.7 & Master's degree & 8 & 44.4 \\
    Other & 1 & 5.6 & Doctoral degree & 1 & 5.6 \\
    Prefer not to say & 8 & 44.4 & Professional (JD, MD) & 1 & 5.6 \\
    \midrule
    \multicolumn{6}{@{}l@{}}{\textbf{Domain of Expertise}} \\
    \midrule
    \multicolumn{4}{@{}l@{}}{\textbf{Computing \& Data}} & \multicolumn{2}{r@{}}{\;11 (57.9\%)} \\
    \multicolumn{6}{@{}l@{}}{\quad\textit{Computer Science, Data Science, Statistics, AI/ML}} \\[0.5ex]
    \multicolumn{4}{@{}l@{}}{\textbf{Health \& Life Sciences}} & \multicolumn{2}{r@{}}{\;\;3 (15.8\%)} \\
    \multicolumn{6}{@{}l@{}}{\quad\textit{Public Health, Healthcare, Bioinformatics}} \\[0.5ex]
    \multicolumn{4}{@{}l@{}}{\textbf{Social \& Behavioral Sciences}} & \multicolumn{2}{r@{}}{\;\;5 (26.3\%)} \\
    \multicolumn{6}{@{}l@{}}{\quad\textit{Social Science, Anthropology, Computational Social Science}} \\
    \bottomrule
    \end{tabular}
    \label{tab:demographics}
\end{table}

\subsection{Participants}

We recruited 18 participants and compensated them \$15 per hour. Consistent with HCI usability research standards \citep{nielsen2006prioritizing}, this sample size enables rich qualitative analysis and identification of critical usability issues, which is appropriate for mix-initiative system design and iterative improvement. This approach provides essential foundational evidence prior to large-scale validation studies. The HCI literature demonstrates that N=18--20 is sufficient for identifying 85\% of usability issues in interactive systems \citep{virzi1992refining, turner2006determining}.


Demographic information is reported in Table \ref{tab:demographics}. Participants were recruited through multiple social media groups and were aged 18 or older, with no restriction on recruitment to our university affiliates. The sample included a broad age range, with 72.2\% under 30. Gender representation was balanced (55.6\% female, 44.4\% male), and participants reflected diverse racial and political backgrounds, with 44.4\% opting not to disclose political affiliation. 

Educational attainment was high, with 44.4\% holding a Master’s degree and 11.2\% holding doctoral or professional degrees. Expertise was concentrated in computing and data-related fields (57.9\%), with additional representation from social and behavioral sciences (26.3\%) and health and life sciences (15.8\%). This sample offers a range of valuable perspectives for a constructive evaluation, spanning both social media researchers and individuals with varying levels of AI experience.

Digital informed consent for participation in the study has been obtained from all participants prior to the study. After that, participants completed a pre-study Qualtrics questionnaire\footnote{\url{https://www.qualtrics.com}} about their research background, OSN familiarity, and prior AI use. For descriptive analysis, we grouped participants by self-reported OSN-research experience: the \textit{OSN-novice} group ($n=9$) had not previously conducted social media research, whereas the \textit{OSN-experienced} group ($n=9$) had used OSN data in academic research. This grouping does not denote AI expertise; most participants (n=17) reported frequent prior use of AI tools. \add{Two participants reported a high school or GED as their highest completed credential. At the time of the study, both were senior undergraduate students who had conducted research using social media data. We use ``researcher'' in the broader sense throughout this paper, and we use ``experienced'' to refer only to prior use of OSN data in academic research, not to degree or professional status.}

\subsection{Tasks via AI4OSN Platform}

After the pre-survey, participants explored \textsc{AI4OSN} in a mixed-methods, task-based session. We assessed perceived usability, workload, and trust-related experiences, as well as descriptive interaction patterns.
Participants completed the following tasks in sequence:

\textit{\textbf{Task 1: Project Creation.}}
Participants logged in to the platform and created a new research project (as shown in Figures 1-2 in the Supplementary Materials). This task evaluated the clarity of onboarding, navigability, and system feedback during initial setup.

\textit{\textbf{Task 2: Query Configuration.}}
Participants configured a data collection query by selecting multiple keywords and specifying a retrieval limit (as shown in Figures \ref{fig:search_keyword} and \ref{fig:search_results}). This task assessed participants’ understanding of search parameters, constraints, and interface affordances.

\textbf{\textit{Task 3: Relevance Labeling.}}
Participants reviewed retrieved social media posts and manually labeled at least 25 posts as \textit{Relevant} or \textit{Non-Relevant} based on their judgment, as shown in Figure \ref{fig:labeling}. They were asked to articulate their labeling criteria to support interpretive analysis. Subsequently, participants were asked to use the AI Labeler function (Figure \ref{fig:ailabel_prompt}) to examine their initial choices. They were free to accept or revise AI suggestions, with their decisions reflected in labeling behavior and agreement metrics.

\textit{\textbf{Task 4: Insight Generation.}}
Participants engaged with AI-generated summaries and analytic outputs (similar to Figure \ref{fig:ai_summary}) and verbally described insights derived from the results. This task elicited accounts of sensemaking support, perceived quality of the outputs, and trust in AI assistance.

For analysis, we first analyzed task-level performance using descriptive statistics. Task completion times were summarized using means and standard deviations, and task success was assessed qualitatively based on successful completion of required steps. 

To characterize interaction after participants saw AI suggestions, we calculated the percentage of initially AI-disagreed manual labels that participants changed (\textit{Shift}) and the percentage of final manual labels that still differed from the AI suggestion (\textit{Gap}). These measures capture calibration and revision behavior rather than establishing expert-validated ground truth, consistent with the study's focus on characterizing participant engagement with AI-generated suggestions.
Perceived cognitive workload was measured after each task using the Raw NASA Task Load Index (Raw TLX) \citep{hart2006nasa}. Participants rated six workload dimensions: mental demand, physical demand, temporal demand, performance, effort, and frustration on a 7-point Likert scale following each task. Performance ratings were reverse-coded prior to analysis. We report task-level descriptive statistics to characterize workload patterns across the workflow and examine how perceived workload changed between manual and AI-assisted interaction stages.

\subsection{Post-study Questionnaires}
Following Task 4, participants completed an AI trust and sensemaking questionnaire assessing their perceived correspondence between AI output and their own interpretation, together with their comfort using the output in research contexts. Agreement with AI-generated summaries was assessed using a three-item agreement questionnaire rated on a 5-point Likert scale. We report item-level descriptive statistics and a composite agreement score computed as the mean across items. Internal consistency was evaluated using Cronbach’s $\alpha$, and a one-sample $t$ test compared the composite score with the neutral midpoint. These measures represent participant perceptions rather than independently verified output quality.

At the end of the study, participants completed the standard 10-item SUS \citep{bangor2008empirical} to evaluate the overall system, which was scored according to standard procedures and summarized as means and standard deviations. Scores were interpreted relative to established usability benchmarks to contextualize perceived usability. We additionally examined participants’ responses to the SUS item measuring anticipated future use to assess perceived adoption and acceptance of the system.

We also collected 5-point self-reported confidence ratings for data collection, labeling, and analysis before and after the session and report paired $t$-tests with Cohen's $d_z$. These immediate self-reports should be interpreted as perceived confidence following exposure to the workflow, not as evidence of durable learning or independently verified skill improvement.

\begin{table}[!tb]
    \centering
    \footnotesize
    \caption{Codebook for thematic analysis, organized by primary themes and subcodes. The table presents theme definitions describing instances in which participants expressed relevant views. \textbf{\#E} indicates the total number of excerpts associated with each theme, and \textbf{\#M} denotes the number of unique participants who mentioned that theme. Example excerpts are also included.}
    
    \begin{tabular}{m{1.4cm} m{5.2cm} m{2.5cm} m{0.15cm} m{0.15cm} m{2.4cm}}
    \toprule
    \multicolumn{1}{c}{\textbf{Theme}} &
    \multicolumn{1}{c}{\textbf{Sub Codes}} &
    \multicolumn{1}{c}{\textbf{Description}} &
    \textbf{\#E} &
    \textbf{\#M} &
    \multicolumn{1}{c}{\textbf{Example}} \\
    \midrule
        Label Criteria & keyword matching, semantic or context-based reasoning, inspection of additional information such as URLs or images, topic familiarity, uncertain & describe how they decide labeling criteria during manual annotation. & 33 & 17 & \textit{``Keyword detection is definitely the first thing''} (P7) \\ 
        
        \midrule
        
        Platform Strengths & choice of OSN, availability of extra information, helpful AI labels, helpful AI summaries, user experience strengths, ease of use & describe aspects of the platform that were helpful or effective for completing tasks. & 66 & 11 & \textit{``It’s really helpful when we really want to conduct the research''}(P5) \\ 

         \midrule
        
        Platform Limitations & issues with AI labels, issues with AI summaries, data collection challenges, difficulty writing prompts, lack of clear model explanations, burden of manual labeling, irrelevant content, trade-offs across OSNs, user experience issues, need to review raw data, no major issues & describe limitations or difficulties encountered while using the platform. & 44 & 18 & \textit{``maybe know more about different models. What can each model do compared to others''}(P2) \\ 
         \midrule
        
        AI Trust Level & high trust ($>90\%$), medium trust ($50\%$--$90\%$), low trust ($<50\%$), no trust, uncertain for professional use & describe their level of trust in AI outputs used for analysis or insight generation. & 21 & 16 & \textit{``I trust the analysis, and I feel comfortable''} (P3) \\ 
         \midrule
        
        Factors Influencing Trust in AI & accuracy of results, perceived AI reliability, trust influenced by others, more inclusive data, richer information presentation, greater workflow familiarity, professional verification & discuss factors that would increase trust in AI-driven analysis tools. & 24 & 14 & \textit{``showing some reasoning behind the AI label''}(P15) \\ 
         \midrule
        
        OSN Related Concerns & model stability and performance, bias, bots or spam, privacy and confidentiality, data access, no concerns & express concerns about using AI in social media research. & 32 & 16 & \textit{``Confidentiality is definitely another concern''} (P1) \\ 
         \midrule
        
        General Feedback & interest in future versions, overall positive perception, accessibility for broad users, need for clearer guidelines, preference for downloadable reports, very positive evaluation & share additional comments or suggestions that do not fit other themes.  & 19 & 11 & \textit{``this website is very good, relatable, and this could ... be used for job searching``} (P6) \\ 
    \bottomrule
    \end{tabular}
    \label{codebook}
\end{table}

\subsection{Debrief Interview}

After the surveys and platform exploration, we conducted a semi-structured interview of approximately 10--15 minutes. The guide asked participants about their relevance criteria, their responses to AI labels and summaries, trust and verification, workflow challenges, and desired future supports. Interviews were digitally recorded, transcribed from Zoom audio\footnote{\url{https://www.zoom.com/}}, manually checked for accuracy, and de-identified before analysis.

We conducted thematic analysis using \citet{braun2006using}'s six-phase approach and a hybrid coding strategy that combined sensitizing concepts from our research questions with inductively developed codes \citep{proudfoot2023inductive}. Three researchers double-coded all 18 transcripts in Dedoose\footnote{\url{https://www.dedoose.com/}} and resolved disagreements through discussion. We report the resulting 47 codes, 236 coded excerpts, and seven themes in Table~\ref{codebook}. The counts indicate the prevalence of coded discussion in this sample; they are descriptive rather than estimates of population prevalence. For example, the theme \textit{Label Criteria} includes 33 excerpts from 17 participants, with a representative excerpt from participant \#7.

We interpreted the qualitative accounts alongside descriptive task-time, workload, and revision data to identify convergent and divergent evidence. This triangulation contextualizes the observed workflow patterns; it does not establish that AI assistance caused changes in workload, task performance, or confidence.
\begin{table}[!htb]
    \centering
    \footnotesize
    \caption{OSN Usage Preferences. Categories are presented in descending order. Counts (n) and Percentages (\%).}
    \begin{tabular}{m{1.5cm}cclcclcc}
    \toprule
        \textbf{Regularly Used Platforms} & \textbf{n} & \textbf{\%} & \textbf{Frequency} & \textbf{n} & \textbf{\%} & \textbf{Primary Use} & \textbf{n} & \textbf{\%} \\ 
        \midrule
        YouTube & 13 & 72.2 & Less than 30 minutes & - & - & Entertainment & 15 & 83.3 \\ 
        LinkedIn & 11 & 61.1 & 30 minutes to 1 hour & 6 & 33.3 & News and Current Events & 13 & 72.2 \\ 
        Instagram & 10 & 55.6 & 1 to 2 hours & 6 & 33.3 & Social Connection & 12 & 66.7 \\ 
        X/Twitter & 5 & 27.8 & 2 to 4 hours & 4 & 22.2 & Information Seeking & 11 & 61.1 \\ 
        Reddit & 4 & 22.2 & More than 4 hours & 2 & 11.1 & Professional or Academic Use & 9 & 50 \\ 
        TikTok & 3 & 16.7 &  &  &  & Political Discussion & 2 & 11.1 \\ 
        Facebook & 1 & 5.6 &  &  &  & Content Creation & - &  -\\ 
        Other & 3 & 16.7 &  &  &  & ~ & ~ & ~ \\ 
        \bottomrule
    \end{tabular}
    \label{OSN_Usage}
\end{table}

\section{Results}

We first discuss pre-survey results on participants' perceptions of OSN and AI use. We then examine participant interactions with the \textsc{AI4OSN} platform and assess cognitive workload across tasks. SUS results for the overall system and participant perceptions of AI-generated content are presented next. Finally, we analyze changes in confidence for conducting OSN research using post-survey data and discuss additional interview feedback through thematic analysis.

\subsection{OSN Preference}

As shown in Table \ref{OSN_Usage}, participants reported frequent engagement with a diverse set of OSNs, with YouTube (72.2\%), LinkedIn (61.1\%), and Instagram (55.6\%) being the most regularly used platforms, reflecting everyday patterns of content consumption and social interaction. In contrast, platforms commonly emphasized in OSN research, such as X (Twitter) and Reddit, were used by smaller subsets of participants (27.8\% and 22.2\%, respectively), suggesting that many participants interact with these platforms primarily as casual users rather than as expert or power users. 

In terms of usage intensity, most participants reported moderate daily engagement with OSNs: 33.3\% spent 30 minutes to 1 hour per day, 33.3\% spent 1 to 2 hours per day, and 11.1\% spent more than 4 hours per day.  These amounts represent substantial daily engagement and indicate that participants were well-experienced with OSN platforms and interactions. The most common uses included entertainment (83.3\%), news and current events (72.2\%), and social connection (66.7\%), followed by information seeking (61.1\%) and professional or academic use (50.0\%). Relatively few participants reported engaging in political discussion or content creation. This pattern indicates that the sample population is familiar with OSNs as sources of information and professional content, which aligns well with this study's goals. It also highlights a clear need for research tools that support systematic analysis of OSN data. An AI-assisted research platform, such as \textsc{AI4OSN}, can therefore build on participants’ existing practices while accelerating research processes and supporting efficient information-seeking through structured, transparent analysis of OSN data.

\begin{table}[!ht]
    \centering
    \footnotesize
    \caption{AI tool usage, perceptions, and preferences. Counts (N) and percentages (\%) are shown.}
    \begin{tabular}{p{6cm}cc |p{4cm}cc}
        \toprule
        \textbf{Category} & \textbf{N} & \textbf{\%} & \textbf{Category} & \textbf{N} & \textbf{\%} \\ 
        \midrule
        \multicolumn{3}{c}{\textbf{Frequency Usage}} & \multicolumn{3}{c}{\textbf{Verification of AI Outputs}} \\
        Never & - & - & Not at all & 1 & 5.6 \\ 
        Rarely (a few times per year) & 1 & 5.6 & Slightly & 2 & 11.1 \\ 
        Occasionally (monthly) & - & - & Moderately & 10 & 55.6 \\ 
        Frequently (weekly) & 4 & 22.2 & Highly & 5 & 27.8 \\ 
        Very frequently (daily or near daily) & 13 & 72.2 & Completely & - & - \\ 
        \midrule
        \multicolumn{3}{c}{\textbf{Types of AI Tools Used}} & \multicolumn{3}{c}{\textbf{Perceived Trustworthiness}} \\ 
        Chat-based AI tools & 17 & 94.4 & Never & - & - \\ 
        Tools for analyzing data or building models & 8 & 44.4 & Rarely & - & - \\ 
        Tools that provide recommendations or help with decisions & 6 & 33.3 & Sometimes & 4 & 22.2 \\ 
        Tools that help label or tag data & 4 & 22.2 & Often & 7 & 38.9 \\ 
        Tools that analyze images or videos  & 3 & 16.7 & Always & 7 & 38.9 \\ 
        Others & 1 & 5.6 &  &  &  \\ 
        \midrule
        \multicolumn{3}{c}{\textbf{Assessment of AI Tool Impact}} & \multicolumn{3}{c}{\textbf{Need for AI Explanations}} \\ 
        Not at all & - & - & Strongly disagree & 1 & 5.6 \\ 
        Slightly & 2 & 11.1 & Disagree & - & - \\ 
        Moderatly & 4 & 22.2 & Neutral & - & - \\ 
        Significantly & 9 & 50 & Agree & 8 & 44.4 \\ 
        Transformatively & 3 & 16.7 & Strongly agree & 9 & 50 \\ 
        \bottomrule
    \end{tabular}
    \label{AI_preference}
\end{table}

\subsection{AI Preference}

In addition to OSN usage, the pre-study survey examined participants’ prior experience with AI-based tools, including usage frequency, types of tools used, and perceptions of AI effectiveness and trust, as summarized in Table \ref{AI_preference}.
Overall, participants reported widespread and frequent use of AI tools, indicating that AI is already embedded in their everyday work practices. A substantial majority of participants (72.2\%) reported using AI-based tools daily or near daily, while an additional 22.2\% reported weekly use. Only one participant (5.6\%) reported rare use, and none reported occasional or no prior use of AI tools. This high level of adoption suggests that participants are generally experienced AI users rather than novices. As a result, this group is well-suited to evaluating AI-assisted workflows, such as AI4OSN, in OSN research contexts.

Participants’ AI experience focused on chat-based text-generation tools. Nearly all participants (94.4\%) reported using systems such as ChatGPT, Claude, or Copilot for drafting, summarizing, and answering questions. Fewer participants reported experience with more specialized tools, including data analysis or model-building systems (44.4\%), recommendation or decision-support tools (33.3\%), and data labeling or tagging tools (22.2\%). Experience with image or video analysis tools was limited to 16.7\%, and only 5.6\% reported using other AI tools. This gap highlights an opportunity for \textsc{AI4OSN}, which focuses on structured data annotation and analysis beyond conversational assistance.

Despite differences in tool familiarity, participants consistently perceived AI as beneficial to their work. Half of the participants reported improvements in workflow efficiency, while 16.7\% described the impact as transformative. The remaining participants reported moderate or slight improvements, and none reported no benefit. These responses indicate that participants view AI as an integral component of their work rather than a supplementary aid. 
Most participants reported always or often trusting AI-generated outputs (77.8\%), while the remaining participants reported sometimes trusting them (22.2\%). Consistent with this cautious stance, most participants (83.3\%) reported frequently verifying or revising AI outputs. One participant reported never verifying AI-generated results. This pattern reflects active human oversight rather than passive reliance.

Additionally, participants expressed strong expectations for transparency and explainability. Nearly all participants (94.4\%) agreed that AI-based analysis tools should explain how conclusions are produced. This finding reinforces the need for human-centered designs that emphasize transparency, accountability, and user control.

\begin{figure}[!htb]
    \centering
  \includegraphics[width=0.6\linewidth]{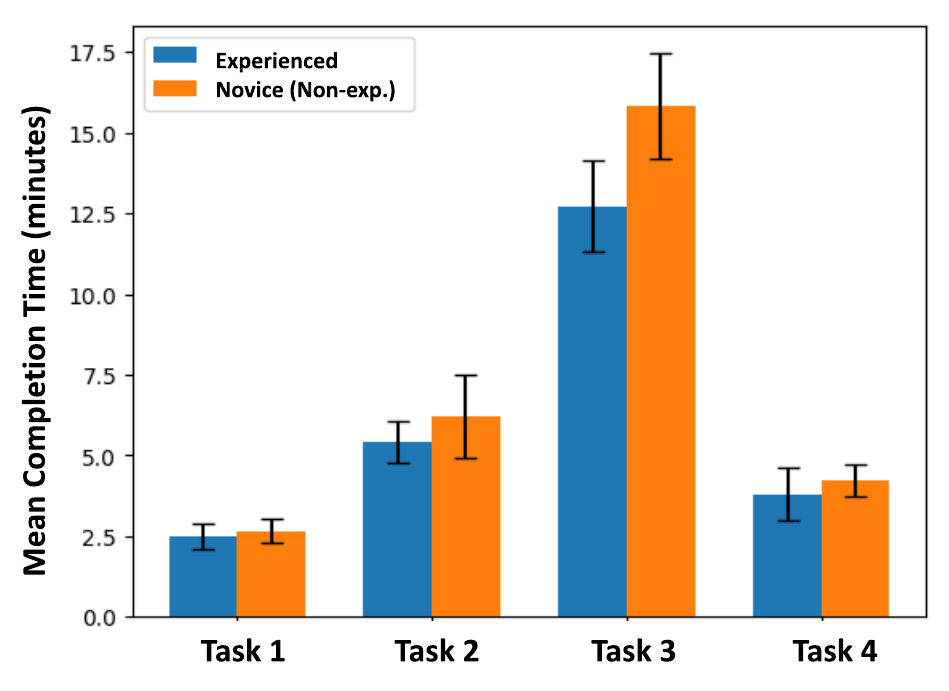}
    \caption{Mean task completion time across four tasks using \textsc{AI4OSN}, with standard error, for two participant groups.}
    \label{fig:time_taken}
\end{figure}

\begin{table}[!htb]
    \footnotesize
    \centering
    \caption{User interactions with \textsc{AI4OSN} for \textbf{Novice Group} (N = 9), including finalized search keywords and exclusion terms, the number of manually labeled posts ($n$), the final prompt used for AI labeling, the percentage (\%) of annotations changed after receiving AI suggestions (Shift), and the percentage (\%) of disagreement between final manual labels and AI suggestions (Gap).}

    \begin{tabular}{m{0.15cm} m{3.2cm} m{1cm}m{0.5cm} m{5.5cm} m{0.5cm} m{0.5cm}}

    \toprule
       \textbf{ID} &
        \textbf{Search Keyword} &
        \textbf{Exclude} &
        \textbf{n} &
        \textbf{Prompt Written} &
        \textbf{Shift} & 
        \textbf{Gap}  \\
    \midrule

    P1 & mental, sport & yoga & 27 & Label each post containing discussion of "sport" and "mental health" otherwise label as unrelated & 25.0 & 33.3 \\ 
        P4 & Lego Smart & - & 25 & Dear AI, please detect whether the tweets are related to the new Lego product that is a smart brick capable of making sounds and lights of different colors and detecting other Lego tiles near it. & 0.0 & 72.0 \\ 
        P5 & Feminist Ethnography History, work & - & 26 & Please label posts that are related to feminist ethnography  & 4.8 & 76.9 \\ 
        P7 & Interview, AI assistance, human-AI collaboration, interviewer, qualitative, conduct interview & job candidate resume & 29 & label each post containing discussion of information seeking oriented interviews as realted, otherview, label as unrelated & 0.0 & 34.5 \\ 
        P9 & Minneapolis & - & 31 & Label posts containing factual reports on what is happening in Minneapolis and people's experiences as related. Otherwise, label people's opinion and jokes about the subject as unrelated & 48.4 & 51.6 \\ 
        P14 & CRICKET NEWS & \textsc{VIOLENCE POLITICS} & 31 & \textsc{CRICKET NEWS ABOUT THE ABOUT PLAYERS AND LEAUES AND INTERNATIONAL MATCHES} & 11.1 & 51.6 \\ 
        P15 & mental health & - & 25 & Label each post containing or encouraging discussion of mental health as Related, otherwise, label as Unrelated & 33.3 & 24.0 \\ 
  P16 & sleep resource, mental health & - & 31 & Label each post containing stress, mental health, and sleep resource & 40.0 & 9.7 \\     
    P17 & sport & - & 26 & label each posts containing discussion of "sport" as Realted, otherwise label as Unrelated & 12.5 & 53.8 \\ 
       & & & & \raggedleft \textbf{Mean} & \textbf{19.5} & \textbf{45.3}\\
    \bottomrule
    \end{tabular}
    \label{user_projects_novice_group}
\end{table}

\begin{table}[!htb]
    \footnotesize
    \centering
    \caption{User interactions with \textsc{AI4OSN} for the \textbf{Experience Group} (N = 9) }
    \begin{tabular}{m{0.15cm} m{3.2cm} m{1cm} m{0.3cm} m{5.5cm} m{0.5cm} m{0.5cm}}

    \toprule
       \textbf{ID} &
        \textbf{Search Keyword} &
        \textbf{Exclude} &
        \textbf{n} &
        \textbf{Prompt Written} &
        \textbf{Shift} & 
        \textbf{Gap}  \\
        \midrule
        P2 & pet, Pet, Dog, dog, Cat, cat, Cute cat, cute cat, Cute dog, cute dog & - & 25 & Label each posts containing discussion of "pet", "dog", "cat" or anything related to people's interest to pets as Related, otherwise, label as unrelated & 0.0 & 60.0 \\ 
        P3 & mental health & - & 28 & label posts containing "funds" as Related, otherwise, label as Unrelated & 10.5 & 60.7 \\ 
        P6 & cloud computing, artificial, machine, analytics, learning & difficult & 26 & label posts containing machine learning relatable stuff & 0.0 & 80.8 \\ 
        P8 & CARS & - & 27 & \textsc{LABEL EACH POST CONTAINING DISCUSSION OF CARS AS RELATED, OTHERWISE LABEL AS UNRELATED} & 0.0 & 92.6 \\ 
        P10 & trump venezuela & election nobel & 26 & Label each posts containing discussion about Trump's invasion of Venezuela and capture of the president Nicolas Maduro as Related, otherwise, label as Unrelated & 18.2 & 34.6 \\ 
        P11 & Hurricane & - & 32 & Please help to label all the posts with "damage" or "loses" or "death" as Related, otherwise, label as Unrealted & 57.1 & 18.8 \\ 
        P12 & MAGA Minnesota & - & 25 & Label posts related to political discussion related to MAGA or what happening in Minnesota. & 15.4 & 44.0 \\
        P13 & golden globes 2026 & - & 50 & If the tweet is related to the golden globes 2026, label it 'related', if not, label 'unrelated' & 0.0 & 86.0 \\ 
         P18 & \textsc{DATE DRIVEN}, glucose prediction, type 2 diabetes & pregnancy  & 27 & No drugs needed, prediction for the patients with 30 mins or 60 mins & 14.3 & 44.4 \\ 
         & & & & \raggedleft \textbf{Mean} & \textbf{12.8} & \textbf{58.0}\\
    \bottomrule
    \end{tabular}
    \label{user_projects_experience_group}
\end{table}

\subsection{User Interactions on AI4OSN}
Task completion times varied across tasks as shown in Figure \ref{fig:time_taken}. Overall, Task 1 had a mean completion time of 2.56 minutes ($SD = 1.12$), and Task 2 had a mean completion time of 5.80 minutes ($SD = 3.00$). Task 3 required substantially more time, with a mean of 14.27 minutes ($SD = 4.72$). Task 4 averaged 4.01 minutes ($SD = 1.98$). 
All participants successfully completed all four tasks using \textsc{AI4OSN}. We present the detailed projects selected by participants when using our platform in Table \ref{user_projects_novice_group} and \ref{user_projects_experience_group}, grouping users into novice and experienced groups based on their self-reported OSN research experience. Overall, participants showed strong interest in \textit{Mental Health and Wellbeing}, \textit{Technology and AI}, \textit{Social and Cultural Topics}, and \textit{Sports and Entertainment}. Many also expressed interest in \textit{Animals and Pets}, \textit{Politics}, and \textit{Hobbies}. 
Among these participants, one-third (P1, P6, P7, P10, P14, P18) used exclusion filters to improve data collection quality. Most did not specify exclusion keywords at the beginning. They added them after reviewing noisy results during initial exploration. 

\begin{figure}[!htb]
    \centering
    \includegraphics[width=0.6\columnwidth]{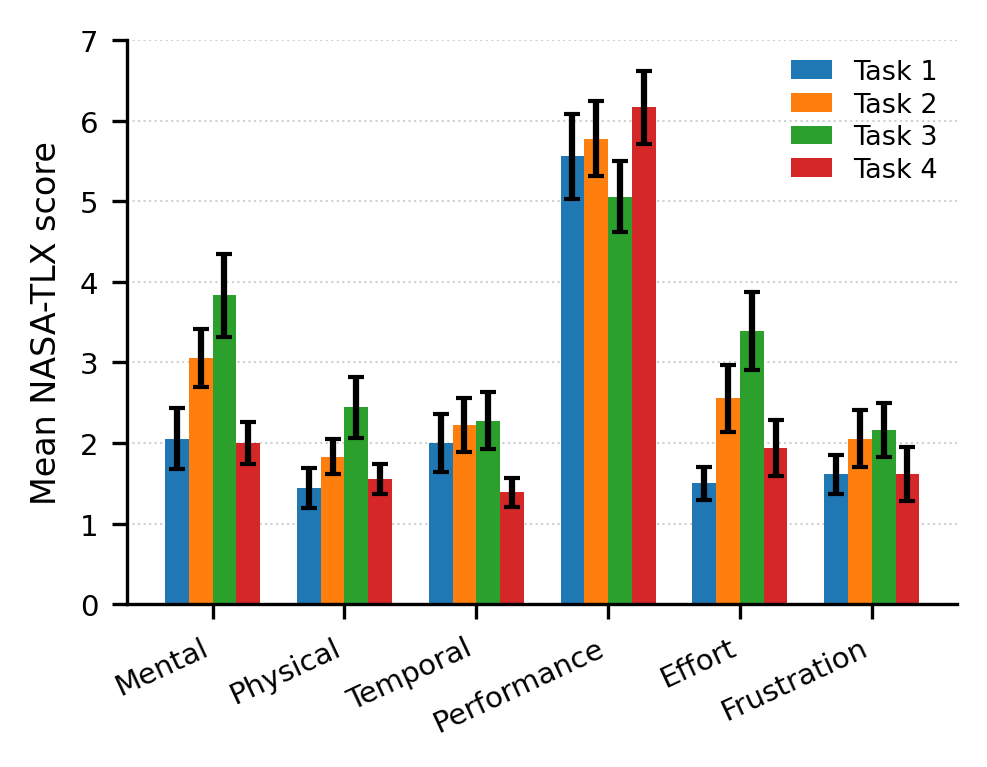}
    \caption{\textbf{NASA-TLX scores across the four task challenges.} Bars represent mean workload ratings for each NASA-TLX dimension, and error bars indicate standard error. }
    \label{fig:tlx}
\end{figure}

Participants collected an average of 52.5 posts ($SD = 19.4$). They manually labeled an average of 28.7 posts ($SD = 5.8$), which constituted the most cognitively demanding activity as reflected in task time (Figure \ref{fig:time_taken}) and NASA-TLX scores (Figure \ref{fig:tlx}). After completing manual annotation, participants explored the AI labeling feature and wrote prompts to guide the model. Most participants wrote a prompt only once, while a few (P2, P15) iterated up to three times to improve AI labeling performance. Most participants used the example prompt template, `Label each post containing a discussion of ``XXX'' as Related; otherwise, label it as Unrelated', either exactly as provided or with only minor revisions. One third of participants included a system message to guide the AI’s overall behavior, for example, ``You are a helper that filters the most relevant posts for future analysis'' (P7) and ``You are a helpful annotator'' (P10). These system messages may help improve the LLM’s ability to generalize to diverse, previously unseen instructions that reflect user values, as discussed in ~\citep{lee2024aligning}. Seven of 18 participants (38.9\%) chose a GPT model over the default GPT-3.5, believing it would yield better results. Five participants (27.8\%) adjusted the model temperature setting, and most of them set it to 0 or near 0 after learning its function, preferring more consistent results. We discuss prompting styles further in Section \ref{discussion_prompt}.

After reviewing AI suggestions, participants changed 16.15\% ($SD=17.88\%$) of the labels on which their initial judgment and the AI suggestion disagreed. Final manual labels still differed from AI suggestions by 51.63\% on average ($SD=23.71\%$); the OSN Novice group showed a smaller final gap than the OSN Experienced group. Rather than establishing which party was more accurate, these values instead highlight a more meaningful pattern: the combination of limited revision and persistent disagreement shows that participants did not simply defer to AI suggestions, and that relevance was interpreted differently across users and topics.

Participants marked an average of 68.7\% of retrieved posts as related ($SD=23.23\%$). This descriptive pattern reflects the participant-chosen queries and criteria rather than the precision of the search mechanism. Together with the interview accounts, it underscores that relevant work requires contextual judgment beyond keyword matching.

\subsection{Cognitive Workload (NASA-TLX)}
\label{section:TLX}
We assessed perceived cognitive workload after each task using the Raw NASA Task Load Index (Raw TLX). Participants rated six workload dimensions, mental demand, physical demand, temporal demand, performance, effort, and frustration, on a 7-point Likert scale following each task. Performance ratings were reverse-coded prior to analysis. Figure~\ref{fig:tlx} summarizes task-wise workload distributions, and we report mean values to highlight trends across tasks.

\textit{\textbf{Task 1: Project Creation.}}
The first task was associated with a low perceived workload across all dimensions. Mean mental demand was low ($M=2.06$), with similarly low physical demand ($M=1.44$) and temporal demand ($M=2.00$). Effort ($M=1.50$) and frustration ($M=1.61$) remained minimal, indicating that participants found project setup straightforward and non-taxing. Reverse-coded performance scores were high ($M=5.56$), suggesting that participants felt successful in completing this task. Overall, Task~1 imposed minimal cognitive burden and served as a gentle onboarding step.

\textbf{\textit{Task 2: Query Configuration.}}
Mean mental demand rose to $M=3.06$, reflecting the cognitive effort required to translate research goals into query parameters. Effort also increased ($M=2.56$), while temporal demand remained moderate ($M=2.22$), indicating that participants did not feel rushed. Frustration remained relatively low ($M=2.06$). Performance ratings remained high ($M=5.78$), suggesting that, despite increased cognitive effort, participants remained confident in their task outcomes.

\textbf{\textit{Task 3: Relevance Labeling.}}
The third task had the highest workload across dimensions, consistent with prior work showing that manual annotation is often the most challenging stage of the OSN research workflow \citep{wang2025lata}. Mean mental demand peaked at $M=3.83$, and effort increased substantially ($M=3.39$), reflecting the sustained attention and judgment required for manual labeling. Physical demand also increased relative to earlier tasks ($M=2.44$), while temporal demand remained moderate ($M=2.28$). Frustration increased modestly ($M=2.17$), suggesting that some participants experienced difficulty during prolonged review. Despite this, performance ratings remained favorable ($M=5.06$), suggesting that participants felt capable but cognitively taxed during this stage.

\textbf{\textit{Task 4: Insight Generation.}}

\remove{Task~4 had lower reported mental demand (\$M=2.00\$) and effort (\$M=1.94\$) compared to the preceding manual-labeling task; temporal demand was \$M=1.39\$, frustration was \$M=1.61\$, and performance was \$M=6.17\$.} \add{Task~4 had lower reported mental demand ($M=2.00$) and effort ($M=1.94$) compared to the preceding manual-labeling task; temporal demand was $M=1.39$, frustration was $M=1.61$, and performance was $M=6.17$.} Our sequential, integrated task design \citep{hornbaek2006current} reflects real-world research workflows where data collection, annotation, and analysis occur naturally in succession. While task order and goal differences prevent the isolation of AI's individual contribution, this design enables assessment of the complete workflow experience, which is our primary research goal. \remove{Participants reported substantial decreases in mental effort and frustration when transitioning from manual annotation to AI-assisted insight generation, suggesting that AI assistance effectively reduced cognitive burden during interpretation while maintaining workflow coherence.} \add{Participants reported lower mental effort and frustration when moving from manual annotation to AI-assisted insight generation.}

\begin{figure}[!tbh]
    \centering
    \begin{subfigure}[t]{0.49\linewidth}
        \centering
        \includegraphics[width=\linewidth]{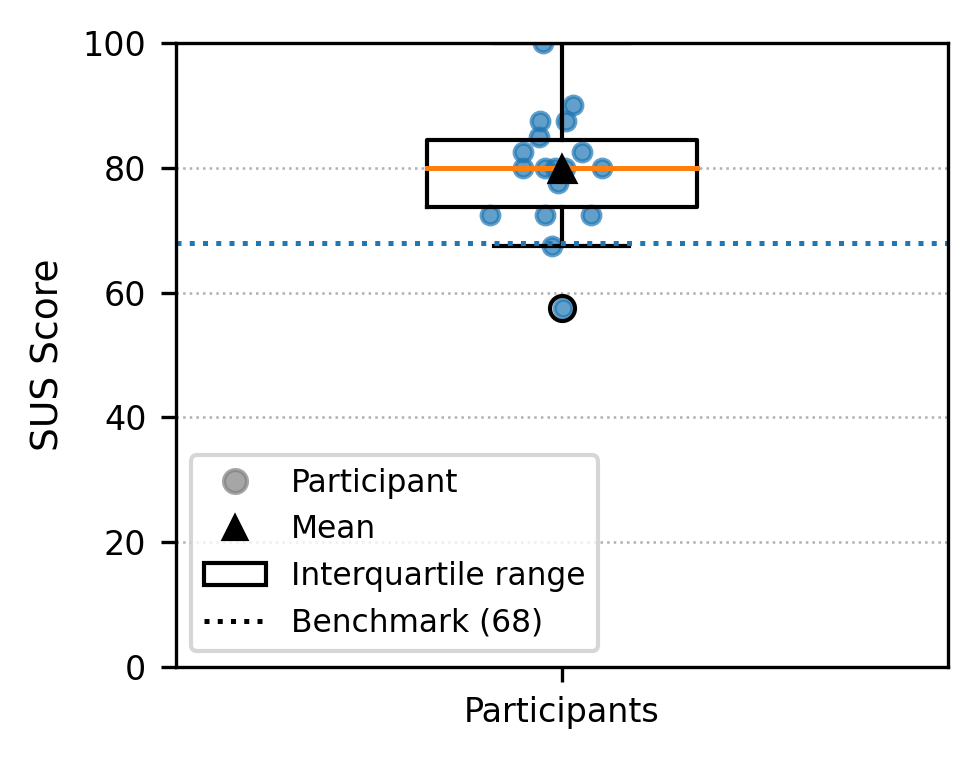}
        \caption{SUS scores with mean, SD, and benchmark baseline}
        \label{fig:sus_individual}
    \end{subfigure}
    \begin{subfigure}[t]{0.49\linewidth}
    
        \centering
        \includegraphics[width=\linewidth]{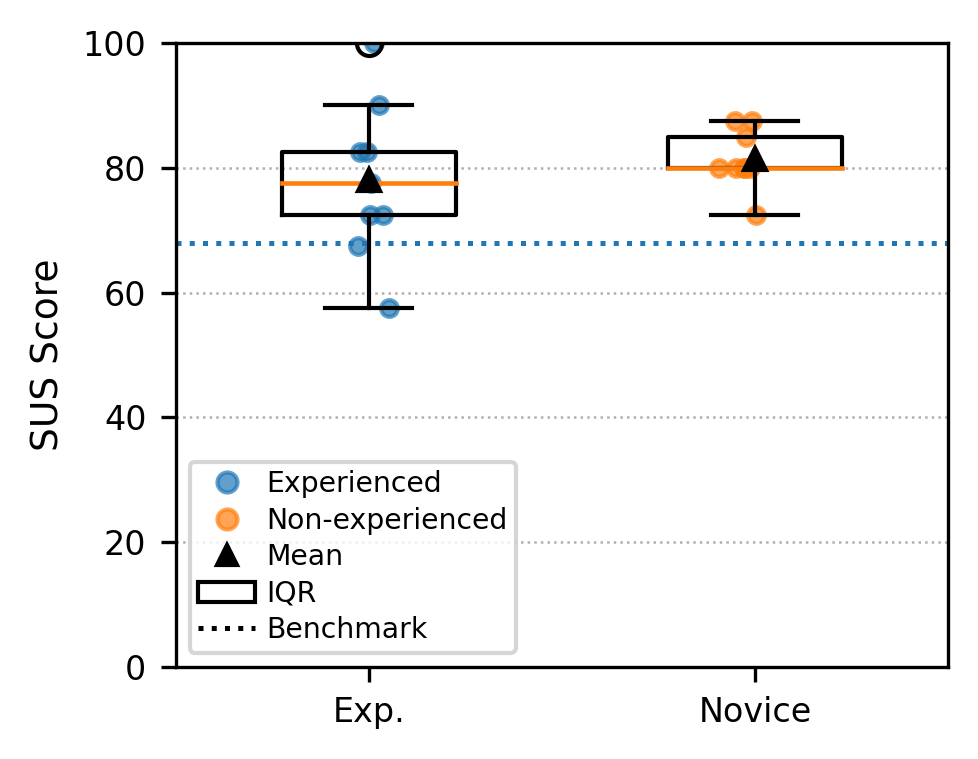}
        \caption{SUS comparison: experienced vs. novice participants}
        \label{fig:sus_group}
    \end{subfigure}
    
    \caption{System usability results. Left: individual participant scores. Right: group-level comparison by prior experience.}
    \label{fig:sus_combined}
\end{figure}

\subsection{System Usability Scale (SUS)}
\label{SUS_results}
We evaluated perceived usability using the System Usability Scale (SUS), a standardized 10-item instrument scored on a 0–100 scale. Across all participants ($N = 18$), the system achieved a mean SUS score of $79.72$ (SD = $9.35$) as shown in Figure \ref{fig:sus_individual}, exceeding the commonly accepted benchmark of 68 \citep{bangor2008empirical}. This places the system in the \textit{above-average to excellent} usability range, with most participants reporting scores well above the benchmark.
Participants also indicated a high likelihood of future use (7-point Likert scale; $M = 5.55$), suggesting that the system is not only usable but also perceived as valuable for continued application.

For descriptive comparison, Figure~\ref{fig:sus_group} reports SUS scores for participants with and without prior OSN research experience. Both groups reported scores above the benchmark (OSN experienced: $M=78.06$; OSN novice: $M=81.39$). A Shapiro–Wilk test confirmed normality. The difference was not statistically significant. These values suggest that positive usability ratings were observed across both groups during this study session, regardless of prior experience.

\begin{figure}[!htb]
    \centering
    \begin{subfigure}[t]{0.49\linewidth}
        \centering
        \includegraphics[width=\linewidth]{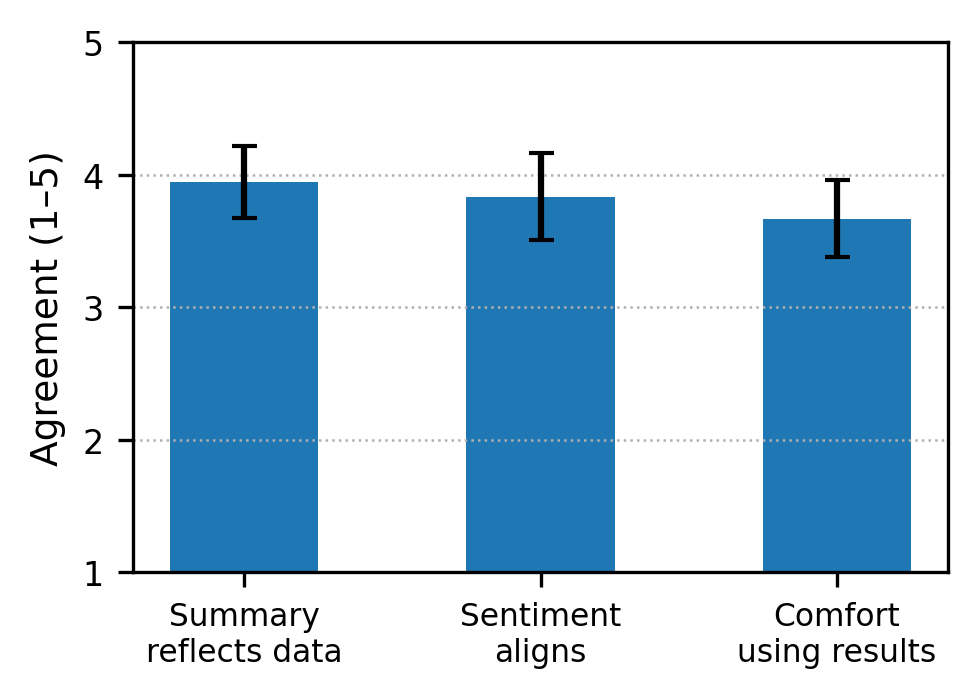}
        \caption{Overall Agreement \remove{with AI with Standard Error}}
        \label{fig:agreement_overall}
    \end{subfigure}
    \hfill
    \begin{subfigure}[t]{0.49\linewidth}
    
        \centering
        \includegraphics[width=\linewidth]{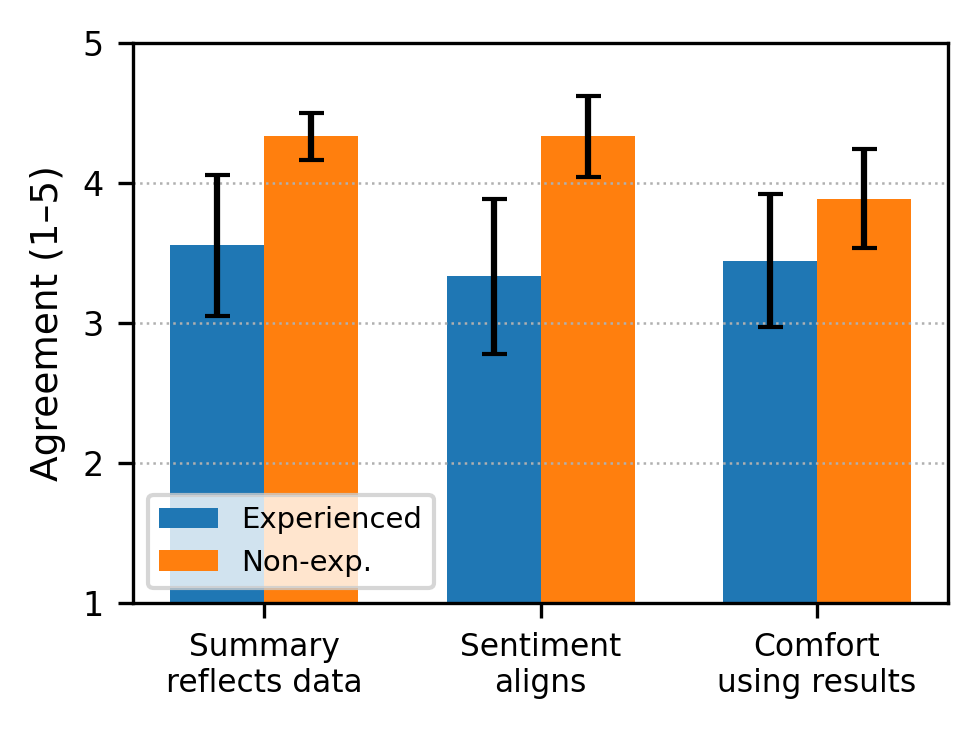}
        \caption{Experienced vs. Novice \remove{Participant} Groups}
        \label{fig:agreement_groups}
    \end{subfigure}
    
    \caption{Overall (left) and group-level (right) participant agreement score with AI \add{and error bars indicate standard error}. Items correspond to three statements: (1)`` The AI-generated summary accurately reflects the data,'' (2) ``The sentiment analysis aligns with my interpretation,'' and (3) ``I would feel comfortable using these results in my research.''}
    \label{fig:agreement_combined}
\end{figure}

\subsection{Agreement with AI-Generated Analysis}

We assessed participants’ agreement with AI-generated analysis content from AI4OSN during Task 4 using three items concerning perceived correspondence with the data, alignment with participants’ interpretations, and comfort using the outputs in research contexts (5-point Likert scale; Figure~\ref{fig:agreement_combined}). Ratings across all items were above the neutral midpoint, indicating generally positive perceptions. Scores were highest for perceived correspondence between the summary and the data ($M=3.94$, $SD=1.16$), followed by alignment with participants’ interpretations of sentiment ($M=3.83$, $SD=1.38$), and comfort with the outputs ($M=3.67$, $SD=1.24$). Between the two groups in Figure~\ref{fig:agreement_groups}, the novice group consistently reports higher agreement with the AI-generated analysis across all items compared to the experience group.

Additionally, we computed a composite agreement score by averaging the three items from Figure~\ref{fig:agreement_combined}, providing an overall measure of agreement with the AI outputs. The composite agreement score was above neutral ($M=3.81$, $SD=1.17$), and the one sample $t$ test differed from the scale midpoint ($t(17)=2.96$, $p=0.009$). The scale showed high internal consistency ($\alpha=0.91$). OSN novice participants reported a higher descriptive mean ($M=4.19$, $SD=0.73$) than OSN experienced participants ($M=3.44$, $SD=1.43$), but this group difference was not statistically significant ($t=-1.38$, $p=0.186$, $d=-0.65$). The small sample and descriptive grouping mean that these values do not establish different trust profiles. Across the sample, participants generally agreed with the generated analysis while still raising concerns about its explanation, reliability, and verification.

\subsection{Pre-post Confidence Gains}
\label{section:pre-post}

\add{We examined changes in participants' self-reported confidence before and after using \textsc{AI4OSN} across three skills: collecting, labeling and annotating, and analyzing data. Confidence was measured on a 5-point Likert scale. Paired $t$-tests were used to assess within-participant changes, with Cohen's $d_z$ reported as the effect size. Shapiro--Wilk tests of the paired difference scores supported approximate normality for most comparisons. Three exceptions were observed: analysis for all participants ($p=.019$), and labeling ($p=.012$) and analysis ($p=.035$) for experienced participants. Wilcoxon signed-rank tests were therefore conducted as sensitivity checks for these comparisons. The substantive conclusions were unchanged: the overall analysis gain remained significant ($W=9.0$, $p=.017$, 12 nonzero pairs), whereas the experienced group's labeling ($W=4.0$, $p=.706$, 4 nonzero pairs) and analysis changes ($W=3.0$, $p=.111$, 6 nonzero pairs) were not statistically significant.}

\begin{figure}[t]
    \centering

    \begin{subfigure}[t]{0.32\linewidth}
        \centering
        \includegraphics[width=\linewidth]{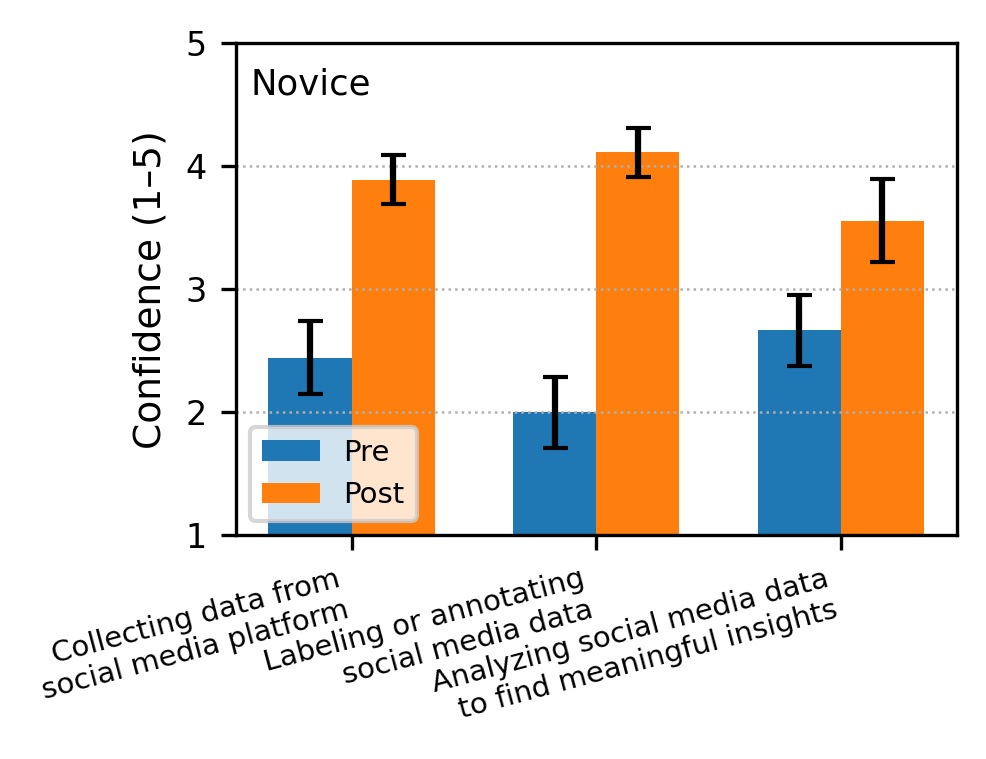}
        \caption{Novice Group}
        \label{fig:conf-novice}
    \end{subfigure}
    \hfill
    \begin{subfigure}[t]{0.32\linewidth}
        \centering
        \includegraphics[width=\linewidth]{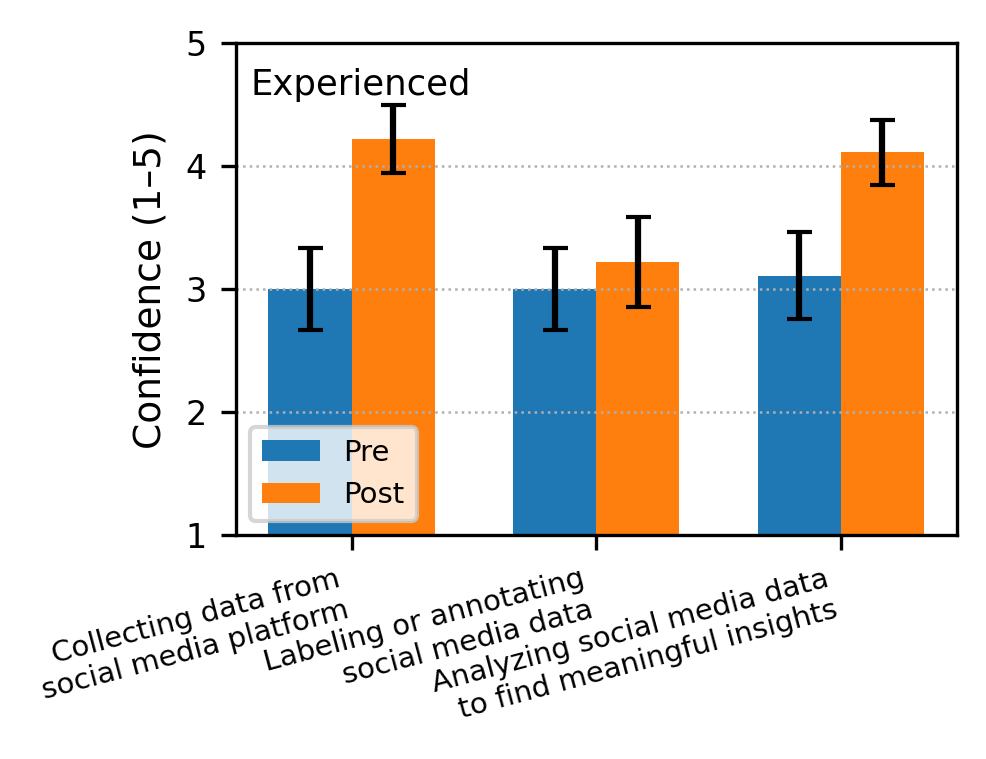}
        \caption{Experienced Group}
        \label{fig:conf-exp}
    \end{subfigure}
    \hfill
    \begin{subfigure}[t]{0.32\linewidth}
        \centering
        \includegraphics[width=\linewidth]{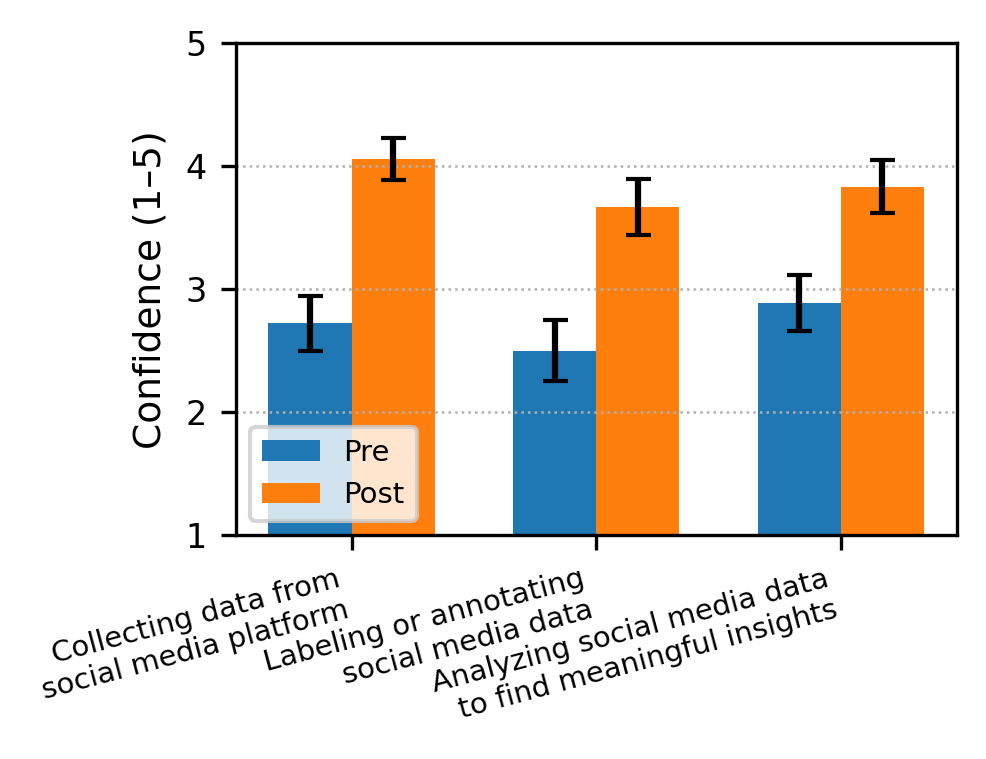}
        \caption{All participants}
        \label{fig:conf-all}
    \end{subfigure}


    \caption{\textbf{Pre–post confidence ratings across participant groups.} Mean confidence scores ($\pm$ standard error) for data collection, annotation, and analysis are reported for novice, experienced, and all participants. \remove{Consistent increases in post-session confidence across all groups indicate improved self-efficacy following system use.} \add{Post-session confidence was higher than pre-session confidence for all participants and for most group-skill combinations. The experienced group's labeling confidence was stable, with a small mean increase.}}
    
    \label{fig:confidence_combined}
\end{figure}

\add{Across all participants ($N=18$), confidence increased significantly for all skills (Figure~\ref{fig:conf-all}). Confidence in data collection increased from $M=2.72$ to $M=4.06$ ($\Delta M=1.33$; $t(17)=4.97$, $p<.001$, $d_z=1.17$), labeling and annotation increased from $M=2.50$ to $M=3.67$ ($\Delta M=1.17$; $t(17)=3.21$, $p=.005$, $d_z=0.76$), and analysis increased from $M=2.89$ to $M=3.83$ ($\Delta M=0.94$; $t(17)=2.72$, $p=.015$, $d_z=0.64$). Under a Holm correction across all nine confidence tests, the collection and labeling gains remain significant (adjusted $p=.001$ and $p=.035$), and the analysis gain is $p=.073$.
We further examined pre--post changes separately for novice and experienced participants (Figures~\ref{fig:conf-novice} and~\ref{fig:conf-exp}). Among experienced participants, confidence in data collection increased significantly from $M=3.00$ to $M=4.22$ ($\Delta M=1.22$; $t(8)=3.05$, $p=.016$, $d_z=1.02$). Labeling and annotation confidence showed a small increase from $M=3.00$ to $M=3.22$ ($\Delta M=0.22$; $t(8)=0.56$, $p=.594$, $d_z=0.18$). Analysis confidence increased from $M=3.11$ to $M=4.11$ ($\Delta M=1.00$; $t(8)=1.73$, $p=.122$, $d_z=0.58$), a moderate trend.}

\add{Among novice participants, confidence increased significantly for data collection, from $M=2.44$ to $M=3.89$ ($\Delta M=1.44$; $t(8)=3.83$, $p=.005$, $d_z=1.28$), and for labeling and annotation, from $M=2.00$ to $M=4.11$ ($\Delta M=2.11$; $t(8)=4.99$, $p=.001$, $d_z=1.66$). Analysis confidence increased from $M=2.67$ to $M=3.56$ ($\Delta M=0.89$; $t(8)=2.10$, $p=.069$, $d_z=0.70$), a moderate trend.}

\add{Finally, Welch's independent-samples $t$-tests compared pre--post change scores between novice and experienced participants. The novice labeling gain was larger than the experienced labeling gain ($t(15.95)=3.24$, $p=.005$). Change scores were similar between groups for data collection ($t(15.94)=0.40$, $p=.692$) and analysis ($t(14.67)=-0.16$, $p=.879$). The subgroup and between-group analyses were exploratory and should be interpreted as hypothesis-generating due to the small sample size.}

\subsection{Qualitative Interview Findings}
Through systematic thematic analysis, we discuss each theme in detail, as shown in Table \ref{codebook}, and illustrate recurring patterns in participants' experiences.

\subsubsection{Labeling Criteria}
\label{Labeling_Criteria}

Participants described relevance labeling as a multi-stage judgment process rather than a simple keyword matching task. Novice participants (P1, P7, P14, P15) and experienced participants (P2, P3, P8, P12) all reported that they first checked whether posts explicitly contained the target term or a closely related phrase. As P7 explained, \textit{``keyword detection is definitely the first thing.''} This suggests that the lexical matching function served as an entry point for topic assessment across users, regardless of experience.

However, participants consistently emphasized that keyword presence alone was insufficient for final decisions. Among experienced participants, this deeper assessment was often described as a more structured, domain-oriented process. P2 looked for \textit{``a bit of context related to the keyword''} while P6, P11, P12, and P13 described using broader thematic alignment, additional conceptual markers, discourse authenticity, and indirect topical references to judge relevance. These accounts suggest that experienced participants tended to treat labeling as an interpretive task that required connecting individual posts to wider semantic or thematic patterns.
Similarly, novice participants also relied on contextual reasoning, though their criteria were more often framed in terms of intuitive, personal, or example-based judgments. P5, P7, P9, P15, and P16 described assessing whether posts reflected the intended topic meaning, excluding jokes or overly opinionated content, distinguishing between close topical variants, and considering whether the content would meaningfully contribute to the discussion. This indicates that they were more likely to ground their decisions in personal familiarity or immediate impressions of relevance rather than explicitly articulated analytical rules.

Interestingly, P5 and P7 incorporated supplementary cues when textual evidence was limited. They manually check URLs, images, or linked content from social media posts for additional context, while P4 relied on prior familiarity with a niche product domain and stated they were \textit{“very confident in assessing whether the tweets were related or not.''} These responses suggest that participants sometimes extended the unit of analysis beyond the post text itself when determining relevance was difficult.

Despite these strategies, ambiguity persisted as a key part of the task. P3 noted uncertainty when posts were only loosely connected to their selected topic, and P4 reported low confidence in AI-generated labels for a niche, emerging topic. Overall, the findings show that participants across both groups combined keyword screening with contextual interpretation and occasional use of external cues, but differed in how they operationalized these strategies. Experienced participants generally applied more structured and domain-driven criteria, whereas novice participants relied more on intuition, personal relevance, and surface-level signals.

\subsubsection{Platform Strengths}
\label{Platform_Pros_Strengths}
Participants described the strengths of \textsc{AI4OSN} as effective AI support, a clear and navigable workflow, and low barriers to use. The AI-generated summary emerged as the most consistently valued feature. 77.8\% (n = 7) of experienced participants 
characterized the summaries as clear, comprehensive, and well aligned with their own interpretation of the data. For example, P3 described the summary as \textit{``super complete\ldots and well structured,''} while P6 noted that it was \textit{``very easy to understand.''} P2 further suggested that the summary offered a more holistic account than manual summarization. Novice participants expressed similar views, with P1, P5, P9, and P15 emphasizing that the summaries helped them quickly grasp narrative and sentiment patterns.  These responses suggest that the summary function supported both interpretability and sensemaking across different levels of expertise.

Participants also viewed AI-assisted labeling as a major strength, though its perceived value differed somewhat by experience level. Experienced participants P6, P8, and P10 emphasized efficiency and alignment, describing the feature as helpful for accelerating decisions and reducing effort. P6, for instance, noted that the AI labeling was \textit{``literally very easy and fast,''} and that the AI-labeled posts closely matched their own judgments. Novice participants, such as P11 and P15, valued this feature more as a reflective aid, particularly on the revision page, where they could compare their own labels with AI outputs side by side. This pattern suggests that experienced users tended to see AI labeling as a time-saving analytical support. In contrast, novice users also used it to calibrate or reconsider their decisions in uncertain cases.

A third recurring strength was the platform's interface and workflow, which many participants linked to the system's overall usability. Most participants (61\%) provided explicit positive feedback on the interface, highlighting the clarity of the layout, visible navigation, and step-by-step task progression. P1 described the workflow as \textit{``very straightforward''} and \textit{``sequential in nature,''} while P9 appreciated features such as sorting posts by recency or engagement. These responses suggest that participants did not view usability as separate from AI support, but rather as a condition that made the platform's AI features easier to interpret and apply.

Finally, accessibility and ease of use were especially salient for participants with less technical confidence. P1 described the platform as \textit{``quite intuitive and very simple to follow,''} P16 emphasized that it was accessible regardless of technical background, and P8 noted that the instructions were very clear. Overall, the findings indicate that participants perceived \textsc{AI4OSN} as strong not only because it provided useful AI assistance, but also because that assistance was embedded in a transparent, accessible, and easy-to-navigate workflow. While both groups converged on the value of summaries and interface clarity, experienced participants more often emphasized AI labeling alignment and efficiency. In contrast, novice participants more often highlighted accessibility and support for sensemaking.

\subsubsection{Platform Limitations}
\label{Platform_Limitations}
AI label reliability was a major discussion across both groups. Experienced participants P8, P10, and P12 emphasized instability, non-determinism, and incorrect relevance judgments, with P10 noting that LLM outputs can vary even at low temperature settings. Novice participants P9, P13, and P17 similarly reported disagreements with AI labels, with P13 stating that the output \textit{``seemed way off what I was expecting.’’} and indicated they would rely entirely on their own judgment. These responses indicate that confidence in AI assistance depended not only on perceived accuracy but also on consistency and predictability.

Participants also pointed to gaps in effort and guidance across the broader workflow. P6 and P9 described manual labeling as burdensome when reviewing posts one by one.  This feedback highlights the potential value of AI-assisted data curation. Prompt writing was especially difficult for P5 and P9, while P2 requested more example prompts, and both P2 and P7 wanted clearer explanations of differences across GPT models. Overall, these findings collectively point to opportunities for improvement in interface clarity, AI label reliability, and prompt-writing guidance.

\subsubsection{AI Trust Level}
\label{AI_Trust_Level}
Participants reported varied levels of trust in AI-generated outputs. Overall, one-third of participants (P2, P3, P6, P8, P14, P15) expressed high trust, describing the AI analysis or summaries as accurate and helpful for completing tasks; for example, one participant stated, \textit{``I'm comfortable using AI labeling because it will help us complete the labeling task, no need to do it manually.'' }(P8). 
Seven participants (38.9\%) reported moderate trust, indicating they were willing to use AI outputs as references but preferred to verify results before fully relying on them, as P11 explained: they were “very willing to take the result as a reference” but would not use it directly. In contrast, a smaller portion of participants (16.7\%, n = 3) expressed low trust, often raising concerns about mislabeled data or uncertainty about the training data, with one participant noting that they were ``not believing them'' and would still label posts manually (P7). 
In contrast, one participant (P4) reported no trust in the AI, while another (P2) expressed uncertainty about using AI outputs in professional or research contexts. Overall, participants tended to calibrate their trust, using AI support cautiously alongside human judgment.

\subsubsection{Factors Influencing Trust in AI}
\label{Factors_Influencing_Trust}
Participants identified several factors that shaped their trust in AI-generated outputs, emphasizing that trust develops through transparency, accuracy, experience, and external validation rather than being assumed by default. All experienced participants explicitly discussed accuracy, reliability, and familiarity with the workflow. They noted that consistent, correct outputs increased confidence over time (P4, P6), and trust was reinforced when AI tools aligned with prior experience or established professional practices (P6, P12). Several experienced users also highlighted the value of quantitative indicators, references, or reasoning in summaries for better understanding of AI outputs (P3, P10, P11, P18). 88.9\% (n=8) of novice participants similarly emphasized the importance of transparency and explanations. They expressed a need for AI labels and summaries to provide reasoning, supporting resources, or context to build confidence in the outputs (P1, P7, P15, P17). Some highlighted additional factors, such as inclusive and representative data (P16) and verification by experts or trusted individuals (P4, P9). Social influence also played a role, with trust increasing when recommendations came from trusted individuals (P4). Overall, participants described trust as developing through transparency, accuracy, experience, and external validation rather than being assumed by default.

\subsubsection{OSN Related Concerns}
Participants reported several concerns about AI-assisted social media analysis tools, with privacy and confidentiality emerging as the most common issues. Two-thirds of participants (66.7\%, n=12) expressed concern that social media data, even when publicly available, could still expose personal beliefs, identities, or other sensitive attributes. As P1 noted, this \textit{``could have harm on people if they’re identified.’’} Similar concerns were raised by P2 and P14, who questioned how AI systems access, learn from, and potentially reuse online data without clear visibility into downstream use. While both novice and experienced participants raised privacy concerns, experienced participants were more likely to situate them alongside broader issues such as data representativeness and system security.

Consistent with prior work \citep{anzum2022biases, wang2023mitigating}, concerns about output quality were also common, with 44.4\% of participants (n=8) raising them. Participants worried that biased training data, skewed post weighting, and the overrepresentation of bot-generated or repetitive content could distort analytical outcomes (P9, P10, P13). Some also questioned whether AI-generated summaries and labels could present inaccurate or misleading interpretations without enough transparency or explanation (P7, P15). Novice participants often framed these concerns in terms of fairness and misrepresentation, whereas experienced participants more often pointed to underlying technical sources of error and instability.

A smaller group of participants focused more specifically on system dependability in practice. P12 and P18 emphasized that inconsistent outputs could undermine trust in real analytic settings. At the same time, a few experienced participants (P3, P6) reported no major concerns. Overall, these findings suggest that privacy concerns were broadly shared, while experienced users tended to articulate more specific technical risks and novice users more often expressed broader apprehension about fairness, misuse, and loss of control.

\subsubsection{General Feedback}
Participants expressed broadly positive perceptions of the system and interest in its future use. Most participants (77.8\%, n=14) described the tool as helpful, accessible, and effective for tasks such as accessing, filtering, and labeling social media data (P3, P5, P10, P13, P18). A recurring strength was the integration of multiple research steps into a single workflow, which participants described as intuitive, efficient, and in some cases \textit{``easy to trust,’’} especially when transparency features such as source links were available (P6, P18).

Positive views appeared across experience levels, though with different emphasis. Experienced participants highlighted reliability, efficiency, and professional utility, often framing the system as a way to reduce technical barriers and support broader analysis. Novice participants also found the system approachable, but more often suggested the need for clearer onboarding, step-by-step guidance, and downloadable reports for offline review (P2, P7, P17).

Interest in future development was also clear. One third of participants (n=6) explicitly suggested extensions such as collaborative interpretation support, broader data sources beyond X/Twitter, and applications in domains like job searching or professional networking (P6, P7). At the same time, some experienced AI users expressed a balanced view, recognizing both the system’s usefulness and the persistent limits of AI, as reflected in P7’s comment that \textit{``the concern will never be as low as 0\%.’’} Overall, the findings suggest strong acceptance of the system, alongside expectations for greater guidance, transparency, and extensibility

\section{Discussion}

This mixed-method study examines how participants used a researcher-controlled AI-assisted workflow across query construction, data collection, relevance labeling, and early-stage sensemaking. We synthesize the findings as evidence about workflow experience, calibration of AI suggestions, and the support participants considered necessary for responsible use.

\subsection{Experiencing a researcher-controlled workflow}

As reflected in the SUS score (Section \ref{SUS_results}) and NASA-TLX rating (Section \ref{section:TLX}), \textsc{AI4OSN} demonstrates strong overall usability, and aligns with participants’ reported likelihood of future use ($M=5.55$), indicating both ease of learning and perceived value for recurring research tasks. \remove{A notable outcome of the study is the magnitude and consistency of pre-post confidence gains across the OSN workflow.} \add{A notable outcome of the study is the pre-post increase in self-reported confidence across the OSN workflow.} Participants’ confidence increased significantly for data collection (the largest gain), labeling, and analysis, \remove{particularly among the novice group, which had little prior exposure to OSN research, as shown in Section \ref{section:pre-post}.} \add{with the novice group showing the larger labeling gain (see Section \ref{section:pre-post}).} \remove{These improvements suggest that \textsc{AI4OSN} may function not only as an automation tool but also as a skill-building scaffold, helping users learn} \add{These self-reported improvements are consistent with a possible role for \textsc{AI4OSN} as a skill-building scaffold, not only an automation tool, although skill was not measured directly. Such a role may involve helping users learn} how to operationalize research goals into queries, structure annotation criteria, and interpret results. This matters for non-technical, non-experienced researchers, who may be capable of substantive interpretation but face barriers to data access and workflow orchestration. These findings further validate the insights from our initial study \cite{wang2025lata}. Additionally, our results suggest that supporting confidence should not mean increasing automation to the point of disengagement. Participants changed only a small portion of labels after AI suggestions, indicating that they remained actively involved and retained interpretive control. In OSN research, where meaning depends on context, norms, and domain expertise, a human-in-the-loop approach such as our AI4OSN may be more appropriate than optimizing solely for maximum agreement with AI.

\subsection{Calibrating AI suggestions against researcher judgment}
\label{discussion_prompt}
Our results highlight a nuanced form of human-AI collaboration in which AI serves as an advisory partner rather than an authoritative annotator. Quantitatively, participants changed relatively few of their initial labeling decisions after seeing AI suggestions (mean shift $=16.15\%$; see Table \ref{user_projects_novice_group}, \ref{user_projects_experience_group}), yet the final mismatch rate between manual and AI labels remained high (mean gap $=51.63\%$). This combination is revealing: participants engaged with AI suggestions but did not substantially defer to them. \remove{Rather than interpreting this gap as a failure of AI labeling, our qualitative data suggest it reflects deliberate calibration: participants most often reported moderate trust,} \add{The gap alone does not show whether the disagreement reflects participant judgment, prompt mismatch, or AI error, and we did not measure accuracy against an independent ground truth. Interview data suggest that participants most often reported moderate trust,} using AI outputs as a reference point while maintaining ownership of labels and verifying discrepancies. This aligns with the pre-study findings that participants commonly verify AI outputs and strongly prefer explainable AI tools. \add{P3's prompt was written around federal mental health grant funding found in the retrieved posts. The keyword ``mental health'' was one of the example keywords in the study instructions, so the P3 gap measures disagreement with the participant's own criterion. P18 wrote a prompt that did not request a relevance label. This was the participant's own prompt without any study intervention as planned. Design recommendations for prompt support are discussed in Section~\ref{sec:prompt-support}.} Participants also reported that AI assistance was beneficial and suggested that AI annotations would be more useful if they included additional context, such as explanations of how the labels were generated, consistent with work showing that appropriate reliance requires support for verification rather than generic transparency \citep{bucinca2021trust, tutul2024trust}. Preliminary interview feedback indicates that many participants found the AI-generated summary helpful for quickly understanding data trends and sentiment without any programming or specialized software skills. The post-task agreement scale further supports this finding. Participants’ composite agreement with the AI-generated analysis was significantly above neutral and showed high internal reliability. Therefore, even though AI labeling did not closely match human judgments, participants still perceived AI outputs as valuable for sensemaking and triage. We found OSN relevance labeling is inherently interpretive and context-dependent. Participants described moving from keyword matching to deeper semantic interpretation, filtering jokes or opinionated content, and sometimes consulting linked resources. These practices likely lead to divergence from an LLM's default interpretation, especially when prompts are brief or underspecified. Indeed, most participants relied on a simple template prompt and did not iterate extensively, which may have limited the model's ability to learn nuanced criteria. Several participants explicitly requested clearer prompt guidance, and many expressed concerns about inconsistent AI labeling.

Participants' prompting behaviors further reveal how researchers engage with AI assistance in structured research tasks. Most participants used the provided template prompt with minimal changes, while fewer introduced system messages or iterated on their outputs multiple times to refine them. When participants adjusted settings (e.g., choosing a stronger model or setting the temperature near zero), they did so with the explicit goal of improving consistency. These behaviors suggest that researchers are developing an emergent prompting literacy that prioritizes stability and reproducibility, values aligned with research practice. However, participants reported difficulty writing effective prompts. They requested more guidance, highlighting the need for interface-level support for prompt development rather than assuming users already possess prompt engineering skills, particularly in a research context.

\subsection{Supports for responsible reliance}
Participants' responses suggest that they view LLM-based assistance as promising for reducing overhead and supporting interpretation, but only when paired with transparency and safeguards. Participants actively downloaded and reviewed the dataset to spot-check AI annotations against their own decisions, reinforcing careful human-in-the-loop validation. They expressed moderate-to-positive agreement with AI-generated analysis and praised the summaries as well-structured and useful for understanding trends. Participants treated AI outputs as reference signals, emphasized verification, and requested explanations, references, and indicators of how the AI arrived at conclusions. \textsc{AI4OSN} enables users to engage more deeply with OSN data and think critically by revising search queries, adjusting prompts, and refining their analyses. This demonstrates that the platform not only supports efficient data exploration but also actively fosters reflective, human-in-the-loop research practices. This points to an important tension in AI-assisted OSN research: LLMs can accelerate analysis, but they can also amplify risks that are already salient in OSN contexts, including privacy concerns, representational harms, and biased or noisy data sources. Indeed, privacy concerns were prevalent in interviews, as were concerns about bias, hallucination, and overconfident system behavior. For OSN research tools, these concerns are not peripheral; they influence whether researchers feel comfortable using AI-generated outputs in real studies. Experienced and novice participants differed in how they calibrated trust in LLM-based systems. Experienced users were more likely to rely on prior experience. \remove{They demonstrated greater accuracy in integrating AI outputs into their workflow, whereas}\add{In contrast,} novice users relied more on transparency, explanations, and external validation to build confidence. This suggests that effective OSN research tools should support both groups by combining reliable performance for experienced users with interpretable and explainable outputs for novices.

\subsection{Design Recommendations for AI-Assisted OSN Research Tools}
Collectively, our results suggest that \textsc{AI4OSN} demonstrates strong usability and adoption potential for end-to-end OSN research (RQ1). The system supports downstream analysis by accelerating sensemaking and offering advisory signals rather than replacing human judgment (RQ2). Our findings also indicate that researchers are more likely to embrace LLM-based assistance when systems foreground transparency, reliability, and responsible handling of OSN data (RQ3). These insights motivate mixed initiative designs that reduce cognitive burden during interpretation while preserving user agency.

\subsubsection{Research-oriented prompt support, not just prompt templates.}
\label{sec:prompt-support}
Instead of treating prompts as an open-ended text box, systems can embed prompting supports that align with research goals: reusable prompt templates linked to projects, versioning to track changes, short ``prompt diagnostics'' (e.g., detecting underspecification), and small calibration sets (e.g., showing a few representative posts and asking the user to confirm criteria). Such features could improve AI label alignment while maintaining transparency and reproducibility. From the user's perspective, systems should provide built-in quality feedback for prompt formulation and reduce the expectation that researchers become prompt engineers. This design direction aligns with recent literature showing that generative AI tools shift work from production to evaluation \citep{simkute2025ironies}, making task allocation, continuous feedback, and interpretability central design concerns.

\subsubsection{Transparent and responsible AI assistance.}
To support responsible adoption, \textsc{AI4OSN}-like systems should incorporate transparency mechanisms tailored to research workflows: (1) clear provenance, including links to underlying posts and the subset of data informing each summary; (2) uncertainty communication, such as confidence cues and ``insufficient evidence'' flags; (3) bias-aware summaries, including indicators of duplicated content, bot-like patterns, or skewed sources; (4) privacy-aware defaults, such as redaction options and reminders for ethical handling of identifiers; and (5) access to raw content, such as original data and AI-generated annotation labels. These features directly address the factors participants identified as critical for trust, including the need for explanations, consistent reliability, and opportunities for external validation.

\subsection{Limitations and Future Work}
\add{This usability study was intentionally scoped for early-stage evaluation. Although the
sample was sufficient to identify key usability patterns and pre--post changes in
self-reported confidence, a larger and more diverse participant population would
increase the precision and generalizability of the findings. In particular, future
studies should recruit sufficient numbers of novice and experienced researchers to
support adequately powered comparisons of how prior expertise shapes interaction with
\textsc{AI4OSN}. The present confidence measures are self-reported and should
therefore be interpreted as complementary to, rather than substitutes for, objective
measures of research performance and annotation quality. A future controlled study
could include a manual baseline, independently validated annotations, and a
counterbalanced task order. Such a design would help distinguish the contributions of
AI assistance, task familiarity, and ordering effects to observed changes in
confidence, annotation quality, and task performance.}

\add{
The current study captures a single session of interaction with the system.
Longer-term deployments are needed to examine how researchers' data-collection
criteria, interpretations, trust, reliance, and expertise develop through repeated
use. A longitudinal design could reveal whether initial confidence gains are
sustained, whether users become better calibrated in evaluating AI outputs, and how
human judgment and AI assistance are integrated into everyday research workflows.
The current evaluation focuses on data collected from X using keyword-based search.
Such queries can introduce sampling bias, potentially amplifying prominent viewpoints
while overlooking less frequent perspectives. Future work should assess
representativeness by comparing retrieved posts with the broader discussion surrounding
a topic. Researchers may also preserve minority viewpoints by selecting a compact
subset of posts that is stratified by sentiment, topic, and emotion before
summarization. The proposed stratified-subset approach was originally developed and
evaluated using product reviews; therefore, its suitability for social media data
should be evaluated for each dataset and research context
\citep{marozzo2025balanced}. Future work should also extend \textsc{AI4OSN} beyond X
to additional platforms and data types, where platform-specific conventions,
moderation practices, and user populations may affect both data quality and analytical
interpretation.}

\add{The AI components in the present study were implemented using GPT-3.5 and GPT-4.
These models provide a useful benchmark for the reported evaluation, but large language
model behavior can change over time and differs across model families. Because
\textsc{AI4OSN} is designed to accommodate newer models, future work should compare
updated models against the present baseline to examine how model choice affects data
collection, labeling, summarization, transparency, and user evaluation outcomes.
Additional directions include developing adaptive agents that respond to users'
expertise and task context, as well as strengthening governance and transparency
mechanisms through auditing tools, provenance records, explanation features, and
interfaces that support users in reviewing and contesting AI-generated outputs. These
directions can help promote more informed, accountable, and human-centered
AI-assisted social media research.}

\section{Conclusion}

We present AI4OSN, a human-centered, multi-agent platform that enables accessible social media and web research. The system integrates LLMs into data collection, annotation, and analysis workflows while preserving human judgment. Our usability study suggests that participants found AI4OSN \remove{improves workflow clarity and reduces cognitive overhead} \add{clarifies workflow steps} for researchers working with complex and noisy social media data. Results indicate that users learn from the dataset through direct interaction with the platform, engaging in iterative refinement of queries, prompts, and analytical strategies. AI assistance supports this process by enabling rapid sensemaking and preliminary analysis without requiring programming or specialized tools. This pattern reflects the value of mixed-initiative design, in which AI augments exploration and understanding rather than replacing human interpretation. Participants reported \remove{significant increases in confidence across all tasks}\add{increases in self-reported confidence for data collection, labeling, and analysis} and benefited from AI-assisted analysis that remained transparent and inspectable. The observed interaction patterns point to design opportunities for agentic AI systems that support user agency, critical reasoning, and responsible research practices. AI4OSN serves as a practical entry point for curating high-quality datasets and generating early insights for downstream analysis. We plan to publicly release the platform framework to support future research.

\section*{DECLARATION OF COMPETING INTEREST}
The authors declare that they have no known competing financial interests or personal relationships that could have appeared to influence the work reported in this paper.

\section*{DATA AVAILABILITY}
Data from this study are not publicly available because of IRB-approved privacy restrictions. All data are stored securely on University of Delaware systems. De-identified data may be shared upon request under a Data Use Agreement (DUA) administered through the University of Delaware Office of Research. Questionnaires and interview materials are provided in the supplementary materials.


\section*{ACKNOWLEDGMENTS}
We acknowledge the development team, Rishi Patel, Aiman Ahmed, Avinash Chouhan, Duc Duy Tran, Joy Mwaria, Ribo Yuan, and Sarah Hershberger, whose work was partially supported by the University of Delaware Undergraduate Research Program and/or the Vertically Integrated Projects (VIP) program. Contributions from Abhin Suresh and Srikanth Kumar Perumalla were supported by the University of Delaware through the National Science Foundation SCIPE's Democratizing Access to Research Software Engineering (DARSE) program, under grant No.~2417814.

\section*{ETHICAL APPROVAL}
All procedures involving human participants in this study were reviewed and approved by the Institutional Review Board (IRB) at the University of Delaware (Protocol \#2037012-3). Digital informed consent for participation in the study has been obtained from all participants prior to the study.

\section*{DECLARATION OF GENERATIVE AI AND AI-ASSISTED TECHNOLOGIES IN THE WRITING PROCESS}
The authors confirm that they have read and agree to comply with the Taylor \& Francis policy on the use of Artificial Intelligence (AI)-assisted technologies. We used ChatGPT, Grammarly, and Overleaf’s autocorrect to refine the text, improve grammar and clarity, and suggest improvements to the article’s flow, organization, and formatting. All outputs were reviewed and revised by the authors, who take full responsibility for the final content.






\bibliographystyle{apalike}
\bibliography{sample-base}

\end{document}